\documentclass[aps,prd,reprint,amsmath,amssymb,longbibliography,
superscriptaddress]{revtex4-2}
\usepackage{graphicx,bm,booktabs,slashed,mathrsfs}
\usepackage{fancyvrb}
\usepackage[colorlinks=true,citecolor=blue,linkcolor=blue,urlcolor=blue]{hyperref}
\newcommand{\dd}{\mathrm{d}}
\newcommand{\ii}{\mathrm{i}}
\newcommand{\tr}{\operatorname{tr}}
\newcommand{\MS}{\overline{\mathrm{MS}}}
\newcommand{\PT}{P^{\mathrm T}}
\newcommand{\PL}{P^{\mathrm L}}
\newcommand{\CF}{C_F}
\newcommand{\UV}{\mathrm{UV}}
\newcommand{\IR}{\mathrm{IR}}
\newcommand{\slashp}[1]{\slashed{#1}}
\newcommand{\br}[1]{\overline{#1}}
\newcommand{\supp}{Supplemental Material}
\begin{document}
\title{Gauge-parameter cancellation and finite form factors in one-loop
quark scattering in the refined Gribov--Zwanziger framework}
\author{Wan Wu}
\email{wuw20phy@gmail.com}
\affiliation{Independent Researcher}
\date{November 26, 2026}
\begin{abstract}
We calculate the one-loop amplitude for distinct-flavor massless quark
scattering in a local BRST-invariant refined Gribov--Zwanziger theory with
minimally coupled quarks. The constraint determinant and two dynamical
quark flavors are included in the action and in the loop sum. Retaining
the refined Gribov--Zwanziger masses in the free field matrix, we combine
the mixed self-energy insertion, complete vertex and box corrections,
and external-leg residues. The difference between a general linear
covariant gauge and a reference gauge vanishes symbolically after removal
of the Stueckelberg regulator in the specified dimensional prescription.
We separate ultraviolet and infrared poles, retaining contributions
hidden by scaleless integrals and by the order of mass and dimensional
limits. Local modified-minimal-subtraction counterterms close the
ultraviolet poles, and the remaining infrared poles cancel in the
stated nonzero-mass domain. The finite answer is supplied as eight
color--spin form factors over a thirty-entry scalar integral inventory,
with all 144 finite-function coefficients provided electronically.
An exact scale-response check closes all eight finite rows against the
running Born term, and a spectral representation isolates a triangle
sector with a regular conditional coalescent-pole limit.
Physical-region benchmarks evaluate every scalar family on a common
continued sheet. A specified small-mass path reproduces four QCD
finite hard helicity coefficients numerically after an explicit IR conversion.
Infrared helicity differences remove the common IR conversion and show
color-dependent mass effects beyond a Born-propagator rescaling.
This revision corrects fourteen coefficients in the Born octet spin
structure following a color-contraction correction; the other seven
rows, gauge cancellation, and exact scale response are retained.
The result defines a partonic amplitude on a specified analytic branch;
it does not assume colored asymptotic states or establish an inclusive
next-to-leading-order cross section.
\end{abstract}
\maketitle

\section{Introduction}
\label{sec:intro}
The restriction of a non-Abelian functional integral in the presence of
Gribov copies modifies the infrared organization of perturbation theory
\cite{Gribov1978,Zwanziger1989}. The refined Gribov--Zwanziger (RGZ)
framework supplements the horizon construction with dimension-two
condensates and yields a gluon two-point function with a nontrivial
infrared mass structure \cite{Dudal2008}. Its formulation in linear
covariant gauges admits a local nilpotent BRST symmetry
\cite{Capri2016} and an algebraic renormalization analysis
\cite{Capri2017}. These developments make it possible to ask a more
restrictive question than the behavior of an individual propagator:
how do the localizing fields, the constraint sector, and the ordinary
quark interactions combine in a complete on-shell four-point amplitude?

A replacement of the gluon propagator inside familiar QCD diagrams is
not, by itself, that calculation. The local action contains mixed
propagators and vertices involving the Stueckelberg and auxiliary fields.
The constraint imposing the transversality of $A^h$ also carries a
functional determinant. Its Grassmann representation must be included
once, together with the corresponding loop contributions. Moreover,
a longitudinal contraction of an individual vertex or box graph produces
terms with a self-energy-like kinematic structure. The pinch technique
provides the standard setting for such rearrangements
\cite{Binosi2009}, but an incomplete set of pinched contributions need
not reproduce the gauge dependence of the complete RGZ amplitude.

Related RGZ calculations have examined the quark propagator at one loop
with minimal matter coupling \cite{deBrito2025}. The present object is
instead the distinct-flavor four-quark amplitude. We retain both
massless flavors in closed quark loops, include the constraint
Grassmann fields in the basic action, and treat the auxiliary-field
matrix in the non-Abelian vertex and self-energy sectors. No fermionic
horizon interaction is added. This specifies the matter theory being
calculated and distinguishes it from other possible RGZ matter
extensions.

Gauge-parameter independence is an expected consistency property of
an appropriately assembled BRST formulation, rather than a new general
BRST theorem. The advance here is the explicit complete four-parton
calculation in this specified local matter theory, including its finite
mass-dependent color--spin structures. The infrared comparison below
distinguishes these structures from a universal exchange rescaling.

Our principal result is an explicit finite integral representation of
that amplitude. We first verify the cancellation of its general gauge
parameter. We then recover the separate ultraviolet (UV) and infrared
(IR) information obscured by scaleless zeros, examine the removal of
the auxiliary regulator, and construct the finite modified-minimal-
subtraction answer. Two color structures and four $D$-dimensional spin
structures generate the result. A scalar representation requires three
$A_0$, twelve $B_0$, nine $C_0$, and six $D_0$ entries at generic
parameters. After known one- and two-point finite parts are extracted,
the eight form factors contain 144 nonzero coefficients multiplying
31 finite building blocks, including the rational unit. These
coefficients are supplied with their conventions in the \supp~\cite{SupplementalMaterial}.

The scope is deliberately partonic. Gauge-parameter cancellation of a
specified amplitude is different from a construction of a nonperturbative
physical Hilbert space. Complex gluon poles also require an analytic
prescription. We therefore retain the branch and kinematic conditions
of the calculation and make no claim about a unitary colored-particle
$S$ matrix or an inclusive hadronic observable. This distinction allows
the explicit four-point result to be used without extending it beyond
its verified domain.

\section{Local theory and conventions}
\label{sec:theory}
\subsection{Fields and action}
We use the Minkowski metric $g_{\mu\nu}=\operatorname{diag}(1,-1,-1,-1)$,
the Fourier convention $\partial_\mu\to-\ii p_\mu$, and
$D=4-2\epsilon$. The fundamental generators satisfy
\begin{equation}
 \tr(T^aT^b)=\tfrac12\delta^{ab},\quad
 [T^a,T^b]=\ii f^{abc}T^c,\quad
 \CF=\frac{N_c^2-1}{2N_c}.
\end{equation}
The implemented gauge group is $SU(3)$, with $N_f=2$. We retain
$N_c$ in color identities and coefficient bookkeeping, setting $N_c=3$
for the implemented physical result. Formal identities in $N_c$ below
do not claim a separately verified arbitrary-$SU(N_c)$ graph sum. The two quark flavors have the same vanishing mass
and interact through
\begin{equation}
 \mathcal L_q=\sum_{f=1}^{2}\ii\bar q_f\gamma^\mu
 (\partial_\mu-\ii g_s A_\mu^aT^a)q_f .
 \label{eq:Lq}
\end{equation}
The quark--gluon vertex is thus $+\ii g_s\gamma^\mu T^a$ before the
external-state ordering sign is applied.

Introduce $h=\exp(\ii g_s\xi^aT^a)$ and the composite field
\begin{equation}
 A_\mu^h=h^\dagger A_\mu h+
 \frac{\ii}{g_s}h^\dagger\partial_\mu h,
 \qquad \partial^\mu A_\mu^h=0.
 \label{eq:Ah}
\end{equation}
The commuting vector fields $\phi_\mu^{ac},\bar\phi_\mu^{ac}$ and
$\beta_\mu^{ac},\bar\beta_\mu^{ac}$ have anticommuting partners
$\omega_\mu^{ac},\bar\omega_\mu^{ac}$ and
$\zeta_\mu^{ac},\bar\zeta_\mu^{ac}$. Their second adjoint index is a
spectator for the covariant derivative. In addition to the
Faddeev--Popov pair $c,\bar c$, the adjoint scalar Grassmann pair
$\eta,\bar\eta$ represents the constraint determinant. Define
\begin{equation}
 \mathscr K_h^{ad}=-\partial^\mu
 \left(\delta^{ad}\partial_\mu+
 g_s f^{abd}A_\mu^{h,b}\right),\qquad
 \mathscr H_h=\mathscr K_h-\mu^2 .
\end{equation}
In particular, expanding $\mathscr H_h$ makes the auxiliary condensate
term explicit:
\begin{equation}
 \mathcal L_{\mu^2}=\mu^2(\bar\omega\omega+\bar\zeta\zeta
                         -\bar\phi\phi-\bar\beta\beta),
 \label{eq:auxmassExplicit}
\end{equation}
with all auxiliary indices contracted. This term is already included
in the compact $\mathscr H_h$ notation, not an additional interaction.
It gives the $p^2-\mu^2$ auxiliary poles and is distinct from the
$M^4$ regulator.
Repeated Lorentz and adjoint indices are contracted. With bars denoting
independent fields in the functional integral, the action used here is
\begin{align}
 \mathcal L={}&-\tfrac14 F_{\mu\nu}^aF^{a\mu\nu}
 +b^a\partial^\mu A_\mu^a+\tfrac\alpha2 b^ab^a
 +\bar c^a\mathscr K^{ad}(A)c^d
 \notag\\
 &+\mathcal L_q \notag\\
 &-\partial^\mu\tau^a A_\mu^{h,a}
 +\tfrac{m^2}{2}A_\mu^{h,a}A^{h,a\mu}
 +\bar\eta^a\mathscr K_h^{ad}\eta^d
 \notag\\
 &+\bar\phi_\mu^{ac}\mathscr H_h^{ad}\phi^{dc\mu}
 -\bar\omega_\mu^{ac}\mathscr H_h^{ad}\omega^{dc\mu}
 \notag\\
 &+\bar\beta_\mu^{ac}\mathscr H_h^{ad}\beta^{dc\mu}
 -\bar\zeta_\mu^{ac}\mathscr H_h^{ad}\zeta^{dc\mu}
 \notag\\
 &+\mathcal L_{\rm hor}+\mathcal L_{\rm reg},
 \label{eq:Ltotal}\\
 \mathcal L_{\rm hor}={}&
 \frac{m_G^2}{2\sqrt{N_c}}f^{abc}A_\mu^{h,a}
 (\phi^{bc\mu}+\bar\phi^{bc\mu}
 -\beta^{bc\mu}-\bar\beta^{bc\mu}),
 \label{eq:Lhor}\\
 \mathcal L_{\rm reg}={}&-\tfrac12 M^4\xi^a\xi^a+\rho\xi^ac^a.
 \label{eq:Lreg}
\end{align}
Here $\mathscr K(A)$ is defined as $\mathscr K_h$ with $A^h$ replaced by
$A$. Integration by parts gives the derivative forms of the ghost
terms. Our mass coordinate is
\begin{equation}
 m_G^4=2N_c g_s^2\gamma^4.
 \label{eq:horizoncoordinate}
\end{equation}
The condensate parameters $m^2,\mu^2$ and the horizon coordinate $m_G^4$
are distinct from the temporary Stueckelberg regulator $M$.

The $\eta$ integral produces $\det\mathscr K_h$ for the transversality
constraint. It is not an extra polarization of an auxiliary vector ghost.
In particular, an $\eta$ loop has no Lorentz trace factor $D$.
Its loops, and the two closed quark loops, are included in the basic
self-energy calculation. Neither is inserted again during
renormalization.

\subsection{BRST symmetry and the regulator}
Let $\mathsf s$ denote the graded BRST differential. The relevant
transformations are
\begin{align}
 \mathsf s A_\mu^a&=-(D_\mu c)^a,&
 \mathsf s c^a&=\tfrac{g_s}{2}f^{abc}c^bc^c,\notag\\
 \mathsf s\bar c^a&=-b^a,&\mathsf s b^a&=0,\notag\\
 \mathsf s h&=-\ii g_s c h,&\mathsf s A_\mu^h&=0,\notag\\
 \mathsf s\phi&=\omega,&\mathsf s\beta&=\omega,\notag\\
 \mathsf s\bar\omega&=\bar\phi+\bar\beta .
 \label{eq:brst}
\end{align}
The remaining auxiliary fields in these multiplets, $\tau$, $\eta$, and
$\bar\eta$ are BRST inert. The quarks transform in the fundamental
representation. For example, invariance of the auxiliary sector follows
from $\mathsf s\mathscr H_h=0$ and the cancellation of
$\bar\phi\mathscr H_h\omega$, $\bar\beta\mathscr H_h\omega$, and the
variation of $-\bar\omega\mathscr H_h\omega$. The horizon term is
invariant because $\mathsf s(\phi-\beta)=0$.

With the source convention of Eq.~\eqref{eq:Lreg},
$\mathsf s\rho=-M^4$ and $\mathsf sM^4=0$ give
$\mathcal L_{\rm reg}=\mathsf s(\rho\xi^a\xi^a/2)$.
At $\rho=0$ but finite $M$, this source slice is not BRST closed:
$\mathsf s\mathcal L_{\rm reg}=M^4\xi^ac^a$.
Consequently a finite-$M$ scattering residual is not an appropriate
criterion for the physical gauge cancellation. We remove the regulator
in the prescription described in Sec.~\ref{sec:limits}, and perform
the gauge comparison at $M=0$. A functional Slavnov--Taylor identity
with all source insertions is not inferred from that process-level
comparison.

\subsection{Mixed propagators and spectral masses}
The complete quadratic form is inverted before expanding in loops:
\begin{equation}
 D_{ij}(p)=\ii[K^{-1}(p)]_{ij}.
 \label{eq:inverse}
\end{equation}
The transverse and longitudinal sectors have dimensions five and eight,
respectively. Their ordered kernels are given in
Appendix~\ref{app:kernels}. In particular,
\begin{align}
 D_{AA,\mu\nu}^{ab}(p)={}&-\ii\delta^{ab}
 \left[f(p^2)\PT_{\mu\nu}(p)
 +\frac{\alpha p_\mu p_\nu}{(p^2)^2+\alpha M^4}\right],
 \label{eq:DAA}\\
 f(z)={}&\frac{z-\mu^2}{\Delta(z)},\qquad
 \notag\\
 \Delta(z)={}&(z-m^2)(z-\mu^2)+m_G^4,
 \label{eq:f}\\
 \PT_{\mu\nu}(p)={}&g_{\mu\nu}-\frac{p_\mu p_\nu}{p^2},\qquad
 \PL_{\mu\nu}(p)=\frac{p_\mu p_\nu}{p^2}.
\end{align}
A transpose of an ordered mixed propagator must be accompanied by
momentum reversal; the Gaussian kernel is not a Hermitian mass matrix.
The spectator color projections used to attach the mixed self-energy
are also retained.

For nondegenerate roots define
\begin{align}
 r_\pm&=\frac{m^2+\mu^2}{2}\pm\ii\Omega,\qquad
 \Omega=\sqrt{m_G^4-\frac{(m^2-\mu^2)^2}{4}},\notag\\
 a_\pm&=\frac{r_\pm-\mu^2}{r_\pm-r_\mp},\qquad
 f(z)=\sum_{\sigma=\pm}\frac{a_\sigma}{z-r_\sigma}.
 \label{eq:poles}
\end{align}
The source square-root branch is fixed, with
$r_+-r_-=2\ii\Omega$. We use this relation, together with
$r_++r_-=m^2+\mu^2$ and $r_+r_-=m^2\mu^2+m_G^4$,
without factorizing complex logarithms.

\section{One-loop construction}
\label{sec:construction}
\subsection{Kinematics and Born term}
The external momenta obey
\begin{equation}
 q(p_1)q'(p_2)\longrightarrow q(p_3)q'(p_4),\quad
 p_i^2=0,\quad p_1+p_2=p_3+p_4.
\end{equation}
We set $s=(p_1+p_2)^2$, $t=(p_1-p_3)^2$, and
$u=-s-t$. The physical scattering region has $s>0$ and $t,u<0$;
exceptional limits $stu=0$ are excluded from the generic formula.
The invariant amplitude is defined by
\begin{equation}
 S_{fi}=\delta_{fi}+\ii(2\pi)^4\delta^{(4)}(P_f-P_i)\mathcal M_{fi}.
\end{equation}
Thus the diagrammatic tree and loop sums represent $\ii\mathcal M$.
Writing $J_{31}^\mu=\bar u(p_3)\gamma^\mu u(p_1)$ and
$J_{42}^\mu=\bar u(p_4)\gamma^\mu u(p_2)$, the tree answer is
\begin{equation}
 \mathcal M^{(0)}=-g_s^2 f(t)
 (T^a)_{i_3i_1}(T^a)_{i_4i_2}
 J_{31}^\mu J_{42,\mu}.
 \label{eq:born}
\end{equation}
The external Dirac equations imply $k_\mu J_{31}^\mu=
 k_\mu J_{42}^\mu=0$ for $k=p_3-p_1$.
The longitudinal tree propagator therefore drops out independently
of any loop cancellation.

\subsection{Completeness and the required truncation}
The expansion is an $\hbar$ expansion with $m^2,\mu^2,m_G^4$ fixed in
the Gaussian kernel, rather than a Taylor expansion in $g_s$ at fixed
$\gamma$. For a connected graph with $E$ external legs and $L$ loops,
\begin{equation}
 \sum_v(n_v-2)=E+2L-2.
 \label{eq:count}
\end{equation}
Quadratic mixing is already resummed in Eq.~\eqref{eq:inverse} and does
not supply additional interaction vertices in this identity.

For a one-loop two-point insertion, Eq.~\eqref{eq:count} permits one
four-leg vertex or two three-leg vertices. For a four-quark amplitude,
the same identity alone would permit six-leg vertices. The actual
action provides the additional restriction: each quark flavor couples
only through a $\bar q q A$ vertex. At least two such vertices are
needed to attach the four external quarks. They consume two units of
Eq.~\eqref{eq:count}; at most two remain for pure gauge or localizing
interactions. Those interactions therefore require at most four legs.
A closed quark loop consumes two further $\bar q q A$ vertices and
belongs to the $AA$ self-energy bubble.

Write $A^h=A^h_{(1)}+A^h_{(2)}+A^h_{(3)}+\cdots$, where the subscript
counts fields. Terms linear in $\tau$ or a horizon auxiliary field
require $A^h$ through three fields; ghost-pair terms require it through
two. The mass term $(A^h)^2$ needs the combinations
$(1,1),(1,2),(1,3),(2,2)$. Consequently the expansion through
$g_s^2$, displayed in Appendix~\ref{app:Ah}, supplies all the required
three- and four-leg vertices. This process-specific counting does
not assert an exact all-order BRST symmetry of a finite field
truncation.

There is one further zero class. Four-leg ghost vertices do exist,
including $\bar\eta\eta A\xi$ and $\bar\eta\eta\xi\xi$.
Their single derivative on the internal antighost produces
$\int\dd^Dq\,q_\mu F(q^2)=0$ when they form a tadpole. This is an
odd-integrand zero in translation-invariant dimensional
regularization, even for a massive auxiliary ghost. It must not be
identified with a massive scaleless tadpole. The nonzero self-energy
classes are thus bosonic tadpoles, bosonic bubbles, ghost bubbles,
and the two quark bubbles. We work about the color- and
Lorentz-invariant zero-field background and omit disconnected vacuum
graphs.

\subsection{Assembly and external residues}
The complete one-loop amplitude is assembled as
\begin{align}
 \mathcal M^{(1)}={}&\mathcal M_{\Sigma}
 +\sum_{\ell=1}^{2}\left(\mathcal M_{\mathrm{AV},\ell}
 +\mathcal M_{\mathrm{NAV},\ell}\right)\notag\\
 &+\mathcal M_{\mathrm{box}}+\mathcal M_{\mathrm{crossed}}
 +2\Sigma_V(0)\mathcal M^{(0)}.
 \label{eq:assembly}
\end{align}
Here $\mathcal M_\Sigma$ attaches the full mixed one-loop self-energy
between two full external boson propagators. AV denotes the Abelian
vertex correction, while NAV contains all admissible localizing-field
connections. The two box routings are retained with their fermion
ordering signs. No separately extracted pinch amplitude is added to
Eq.~\eqref{eq:assembly}.

Each of the four massless external quarks contributes one half of the
on-shell residue correction. In the self-energy convention in which
the raw two-point graph equals $-\ii\Sigma$, the transverse Landau
contribution gives
\begin{equation}
 \Sigma_V(0)=\frac{g_s^2\CF}{16\pi^2}
 \left[-\frac{(D-1)(D-4)}{D}
 \sum_{\sigma=\pm}a_\sigma\frac{A_0(r_\sigma)}{r_\sigma}\right].
 \label{eq:residue}
\end{equation}
Keeping $D$ before the pole expansion is essential. For nonzero
$r_\pm$, the UV residue of $A_0(r_\sigma)/r_\sigma$ is unity and
$a_++a_-=1$; Eq.~\eqref{eq:residue} has the finite limit
$3g_s^2\CF/(32\pi^2)$ at this order. Setting $D=4$ in its prefactor
before expansion would incorrectly discard that contribution.

\subsection{Diagram classes and a local Ward check}
The organization of Eq.~\eqref{eq:assembly} is summarized in
Table~\ref{tab:contributionaudit}. A matrix connection is an ordered
field assignment, not an additional topological loop. In particular,
the two-point vertices used by the diagram generator on bosonic lines
represent the full Gaussian covariance. They are not interaction
insertions to be counted a second time. Ghost and quark lines instead
use their direct free propagators.

\begin{table*}[t]
\caption{Contributions to the distinct-flavor amplitude. The table
specifies algebraic sectors, not a universal number of particle-level
graphs. The full mixed sector includes all ordered matrix entries.
All symmetry factors and closed-Grassmann-loop signs belong to the
generated amputated diagrams.}
\label{tab:contributionaudit}
\begin{ruledtabular}
\begin{tabular}{p{.18\textwidth}p{.48\textwidth}p{.24\textwidth}}
Sector & Required sum & Counting convention\\
\hline
Mixed self-energy & Bosonic tadpoles and bubbles; four ghost species; two quark flavors
 & $(D_0\Pi D_0)_{AA}$\\
Abelian vertices & One correction on each quark line & Two external-line placements\\
Non-Abelian vertices & Full admissible mixed-field connections on each line
 & No additional pinch amplitude\\
Boxes & Direct and crossed routings & Original fermion ordering\\
External residues & Four factors $\delta Z_q^{\rm OS}/2$
 & $2\Sigma_V(0)\mathcal M^{(0)}$\\
Ghost four-point tadpoles & Included before applying their zero test
 & Odd loop momentum, not a missing topology\\
\end{tabular}
\end{ruledtabular}
\end{table*}

A useful independent check of the mixed external legs follows directly
from the multiplier equation. At quadratic order
$A_\mu^{h,a}=A_\mu^a-\partial_\mu\xi^a$; hence its normalized
longitudinal component is $A_L+\ii\sqrt z\,\xi$.
The $\tau$ row of $K_L$ in Appendix~\ref{app:kernels} gives
\begin{equation}
 (e_A^{\rm t}+\ii\sqrt z\,e_\xi^{\rm t})D_L
 =-\frac{e_\tau^{\rm t}}{\sqrt z}.
 \label{eq:linearAhWard}
\end{equation}
Here $e_X$ is the unit vector of field $X$ in the stated ordered basis.
For every insertion column other than $\tau$, the longitudinal
$A^h$ combination is therefore zero. The nonzero $\tau$ entry is the
multiplier contact term and must not be discarded. Equation
\eqref{eq:linearAhWard} has been checked against the inverse quadratic
operator extracted from the action, including all eight columns.
It fixes a mixed-propagator convention; it is not by itself the
one-loop Slavnov--Taylor identity.

For a separate elementary fermion check, define
$\mathcal G_0(k)=1/\slashp{k}$ with its overall diagrammatic $\ii$
stripped. The Ward contraction obeys
\begin{equation}
 \mathcal G_0(k+p)\slashp p\,\mathcal G_0(k)
 =\mathcal G_0(k)-\mathcal G_0(k+p).
 \label{eq:telescoping}
\end{equation}
This identity explains the pole-removing terms in the vertex and box
sum, while Eq.~\eqref{eq:linearAhWard} checks the bosonic mixing
convention. Neither identity permits deleting the unpinched part of a
vertex. The full symbolic sum, rather than a selected set of Ward
contractions, is used in Eq.~\eqref{eq:gaugedifference}.

\subsection{Gauge-parameter cancellation}
At $M=0$ we evaluate
\begin{equation}
 \delta_\alpha\mathcal M^{(1)}=
 \mathcal M^{(1)}(\alpha)-\mathcal M^{(1)}(1)
 \label{eq:gaugedifference}
\end{equation}
with $\alpha$ symbolic. Differences of the six complete vertex/box
integrands are formed before pole decomposition and tensor reduction.
Their sum is then combined with the mixed self-energy and external-
residue differences. This ordering exposes Ward cancellations without
requiring two separately expanded four-point amplitudes.

For illustration, the inverse-quark-propagator identity
\begin{equation}
 \slashp{q}=S_0^{-1}(r)-S_0^{-1}(r-q)
 \label{eq:ward}
\end{equation}
removes an adjacent internal quark pole when applied to an on-shell
external spinor. Both sides of a non-Abelian vertex must be contracted;
applying the right contraction first and then the left contraction to
the remainder counts a double-longitudinal term once. The residual
vertex tensor still contains its internal quark pole. This is why
replacing the complete NAV by only its pinched part would not give
Eq.~\eqref{eq:assembly}.

Using $D$-dimensional Clifford algebra, momentum conservation and the
massless external Dirac equations, all sixteen collected color--spin--
scalar coefficients of Eq.~\eqref{eq:gaugedifference} vanish exactly:
\begin{equation}
 \delta_\alpha\mathcal M^{(1)}=0.
 \label{eq:gaugezero}
\end{equation}
No numerical parameter assignment is used. The sixteen entries refer
to the comparison representation, not a minimal set of physical
helicity amplitudes. Their separate vanishing suffices for
Eq.~\eqref{eq:gaugezero}, without requiring an independence proof of
the spin generators. The tree term and the loop term are separately
gauge-parameter independent.

The color algebra uses the corrected FeynCalc structure-constant
contraction rules~\cite{FeynCalcSUNFFix}, with the relevant contractions
checked independently against SU($N_c$) identities. The software snapshot,
validation identities and changes relative to the earlier version are
documented in the Supplemental Material. All coefficients and numerical
results reported here use the corrected calculation.

\section{Scalar form factors and analytic prescription}
\label{sec:formfactors}
\subsection{Color and spin generators}
Let $C_d=\delta_{i_3i_1}\delta_{i_4i_2}$ and
$C_x=\delta_{i_3i_2}\delta_{i_4i_1}$. We use the singlet/octet basis
\begin{equation}
 C_1=C_d,\qquad C_8=\tfrac12(C_x-C_d/N_c)
 =(T^a)_{i_3i_1}(T^a)_{i_4i_2}.
 \label{eq:color}
\end{equation}
If the color-flow coefficients are $F_d,F_x$, then
$F_1=F_d+F_x/N_c$ and $F_8=2F_x$.
For $B_{31}[X]=\bar u(p_3)Xu(p_1)$ and
$B_{42}[Y]=\bar u(p_4)Yu(p_2)$, define
\begin{align}
 S_1&=B_{31}[\gamma_\mu]B_{42}[\gamma^\mu],\notag\\
 S_2&=B_{31}[\slashp{p_2}]B_{42}[\slashp{p_1}],\notag\\
 S_3&=B_{31}[\gamma_\mu\gamma_\nu\gamma_\rho]
 B_{42}[\gamma^\mu\gamma^\nu\gamma^\rho],\notag\\
 S_4&=B_{31}[\gamma_\mu\gamma_\nu\slashp{p_2}]
 B_{42}[\gamma^\mu\gamma^\nu\slashp{p_1}].
 \label{eq:spin}
\end{align}
They are retained as $D$-dimensional generators; no four-dimensional
Fierz relation is imposed before pole cancellation. The result reads
\begin{equation}
 \mathcal M_{\rm ren}=\mathcal M^{(0)}+
 \sum_{A=1,8}\sum_{r=1}^{4}C_A S_r F_{Ar}^{(1),\rm fin}
 +O(\hbar^2),
 \label{eq:final}
\end{equation}
where each displayed one-loop form factor already contains $g_s^4$
and its loop normalization. The electronic ColorID 2 denotes $C_8$,
not $C_x$.

\subsection{Integral conventions and finite building blocks}
We follow the scalar Passarino--Veltman conventions
\cite{Passarino1979,Denner2011}. For chords $Q_0=0,Q_i$, define
\begin{equation}
 I_n=\frac{(2\pi\mu_R)^{4-D}}{\ii\pi^2}
 \int\dd^Dq\prod_{j=0}^{n-1}
 \frac{1}{(q+Q_j)^2-r_j+\ii0}.
 \label{eq:PVdefinition}
\end{equation}
Its one-, two-, three-, and four-point scalar cases are denoted by
$A_0,B_0,C_0,D_0$. The factors $\ii\pi^2$ produced by tensor
integration and $(2\pi)^{-4}$ from the diagrammatic measure are
included once in Eq.~\eqref{eq:final}. The remaining dimensional
measure conversion is treated in Sec.~\ref{sec:renormalization}.

Scalar integrals are invariant under a relabeling of all denominators,
including a change of zero chord. We use these symmetries only for
scalar functions. Moving the zero chord of a tensor coefficient also
mixes its Lorentz structures and cannot be performed as a scalar
argument permutation. After exact $D$-dimensional tensor reduction,
the scalar inventory is
\begin{equation}
 (n_A,n_B,n_C,n_D)=(3,12,9,6).
 \label{eq:inventory}
\end{equation}
In particular, four-point topology does not imply four distinct
integral functions: different masses and invariant arguments remain
different entries. We do not identify Eq.~\eqref{eq:inventory} with
a minimal four-dimensional helicity basis.

For a scalar integral, write its finite part after common UV/IR
bar-epsilon subtraction as $\bar I_n$. For example,
\begin{align}
 \bar A_0(r)&=r[1-\ln(r/\mu_R^2)],\notag\\
 \bar B_0(v;0,0)&=2-\ln[(-v-\ii0)/\mu_R^2].
 \label{eq:knownfinite}
\end{align}
The other finite functions are retained as precisely defined scalar
integrals, rather than numerically evaluated or implicitly replaced
by a real-mass formula. Complex-mass scalar integral representations
are discussed in Ref.~\cite{Denner2011}.

The final coefficient representation uses
\begin{equation}
 F_{Ar}^{(1),\rm fin}=P_{Ar}
 \sum_{j=0}^{30}c_{Ar,j}\,\mathcal F_j,\qquad
 \mathcal F_0=1.
 \label{eq:coefficients}
\end{equation}
Besides the rational unit, the $\mathcal F_j$ comprise six explicit
logarithms and 24 finite $B_0,C_0,D_0$ functions. The prefactors and
nonzero entry counts are given in Table~\ref{tab:prefactors}; the
complete function catalogue and all coefficients $c_{Ar,j}$ are in
the \supp. The rational coefficients depend on
$(s,t,r_-,r_+,\mu^2,N_c)$, with scale dependence carried by the finite
functions. No additional integral identity is assumed in taking a
common rational factor out of a row.

\begin{table*}[t]
\caption{Exact row prefactors in Eq.~\eqref{eq:coefficients}. Electronic
ColorID 1 is the singlet and ColorID 2 is the octet. Entries count
nonzero finite-function coefficients, including a rational entry when
present; they do not count independent spin or scalar master integrals.}
\label{tab:prefactors}
\begin{ruledtabular}
\begin{tabular}{cccc}
ColorID & SpinID & Entries & $P_{Ar}$\\
1 & 1 & 15 & $\frac{\left(N_c-1\right) \left(N_c+1\right) g_s^4}{16 \pi ^2 N_c^2 (s+t)}$ \\ 
2 & 1 & 31 & $\frac{g_s^4}{4 \pi ^2}$ \\ 
1 & 2 & 19 & $\frac{\left(N_c-1\right) \left(N_c+1\right) g_s^4}{8 \pi ^2 N_c^2 (s+t)}$ \\ 
2 & 2 & 19 & $\frac{g_s^4}{2 \pi ^2 t N_c (s+t)}$ \\ 
1 & 3 & 13 & $\frac{\left(N_c-1\right) \left(N_c+1\right) g_s^4}{64 \pi ^2 N_c^2 \left(r_--r_+\right)}$ \\ 
2 & 3 & 13 & $\frac{g_s^4}{32 \pi ^2 N_c \left(r_--r_+\right)}$ \\ 
1 & 4 & 15 & $\frac{\left(N_c-1\right) \left(N_c+1\right) g_s^4}{32 \pi ^2 s N_c^2 (s+t)}$ \\ 
2 & 4 & 19 & $\frac{g_s^4}{16 \pi ^2 s (s+t)}$ \\ 

\end{tabular}
\end{ruledtabular}
\end{table*}

One compact part of the answer can be displayed without introducing
the full coefficient table. In the singlet $S_1$ row, the two explicit
hard logarithms occur as
\begin{align}
 F_{11}^{(1),\rm fin}\big|_{\ell_s,\ell_u}
 ={}&\frac{g_s^4(N_c^2-1)s}{16\pi^2N_c^2t(s+t)}
 \notag\\
 &\times\left[\ln\frac{-s-\ii0}{\mu_R^2}
       -\ln\frac{-u-\ii0}{\mu_R^2}\right].
 \label{eq:hardlogs}
\end{align}
The remaining terms include mass-dependent triangle and box finite
parts. Equation~\eqref{eq:hardlogs} refers to the explicitly separated
logarithms in Eq.~\eqref{eq:coefficients}; it is not a claim that the
finite scalar functions carry no further logarithms of the invariants.
For example, within the same row the coefficient multiplying
$\bar C_0(0,t,0;0,r_-,r_-)$ inside the prefactor $P_{11}$ is
\begin{equation}
 -\frac{(\mu^2-r_-)^2(2r_--t)(s^2+t^2)}
 {2(r_--r_+)^2s(s+t)}.
 \label{eq:representativeC}
\end{equation}
The corresponding mixed-mass coefficient is
\begin{equation}
 \frac{(\mu^2-r_-)(\mu^2-r_+)(r_-+r_+-t)(s^2+t^2)}
 {(r_--r_+)^2s(s+t)}.
 \label{eq:representativeCmixed}
\end{equation}
These expressions exhibit explicitly how the pole residues and
mass separation enter the finite four-point result.

\subsection{Analytic domain}
The scalar functions are defined by continuation from their
Euclidean parametric domain, with the source branch in
Eq.~\eqref{eq:poles} retained. The nonzero internal squared masses
have positive real parts, the slopes away from massless parameter
faces are nonzero with positive real parts in the endpoint analysis,
and $s,t,u$ remain nonexceptional. For example,
$\operatorname{Re}r_\pm>0$ and
$\operatorname{Re}(r_\pm-t)>0$ are satisfied for positive RGZ mass
parameters and spacelike $t$ on the nondegenerate branch.
The finite functions are continued on the same analytic sheet; a
continuation path crossing a Landau singularity requires a separate
boundary analysis.

This prescription is an integral representation of the calculated
partonic function. We do not assign an unrestricted real-energy
contour or a global numerical continuation to the conjugate RGZ
poles. In particular, conjugation of individual masses cannot be
used to discard an imaginary part of a physical-channel integral.
The nonzero-mass domain, the pole branch, and the invariant argument
order are included in the electronic result.

\section{Regulator limits, UV/IR separation, and renormalization}
\label{sec:limits}
\subsection{Scaleless information and small-mass regions}
A dimensionally regulated scaleless zero does not mean that both
its UV and IR residues vanish separately. Schematically,
\begin{equation}
 B_0(0;0,0)=\Delta_{\UV}-\Delta_{\IR}=0
 \label{eq:scaleless}
\end{equation}
when the dimensional regulators are identified. We retain a ledger
of such terms at the original tensor-integral and scalar-reduction
boundaries. Their coefficients are taken from the actual diagrams,
not inferred by imposing a desired cancellation of the total IR pole.

A simple example illustrates the issue. At nonzero squared mass $a$,
\begin{equation}
 C_0(0,0,0;0,a,0)=\frac{A_0(a)}{a^2}
 -\frac{B_0(0;0,0)}{a}.
 \label{eq:ledgerexample}
\end{equation}
Dropping the second term before distinguishing the two residues
would assign the pole of the first term entirely to the wrong
origin. The unseparated integral identity can remain correct while
the separate UV/IR labels become incorrect.

The regulator introduces small squared masses proportional to
$M^2$. Removing a tadpole at generic $D$ and then expanding around
four dimensions need not agree term by term with the reverse order.
We expand the original scalar coefficients in $e_M=M^2>0$, verify
the bound on their negative powers independently of the saved mass
series, and combine regulator tadpoles with the soft regions of
bubbles. The pole, logarithmic and $\epsilon$-times-pole pieces of
these regions must be combined before the limit is interpreted.
For the present fixed Landau task the general-parameter region
corrections cancel. A graph-weighted check of the full mixed source
also gives zero net unapplied UV/IR transfer. These two facts supply
the missing information that cannot be reconstructed from a
scaleless final zero alone.

\subsection{IR endpoint analysis}
The complex-mass IR classification is made before assigning finite
parts. For a parameter integral with $n$ denominators, $2j$ metric
indices, and moment powers $n_i$, the radial weight away from a
massless two-parameter face is proportional to
\begin{equation}
 \lambda^{-1+j+\sum_{i\in\mathrm{complement}}n_i-\epsilon}.
\end{equation}
Positive powers suppress that collinear endpoint. For the remaining
triangle cases the residue reduces to a one-dimensional moment,
\begin{equation}
 \frac{(-1)^R}{\epsilon_{\IR}}
 \int_0^1\frac{y^a(1-y)^b}{A+By}\,\dd y .
 \label{eq:endpoint}
\end{equation}
Here $R$ is the tensor rank. The condition
$\operatorname{Re}A,\operatorname{Re}(A+B)>0$ fixes the elementary
logarithmic antiderivative. For instance,
\begin{equation}
 C_0(0,0,t;0,0,z)\big|_{\IR}=
 \frac{\ln z-\ln(z-t)}{t\epsilon_{\IR}}.
\end{equation}
We retain a difference of logarithms rather than combining their
arguments across a possible complex branch cut.

After the scaleless and mass-region information is restored, the
IR single and double pole coefficients cancel in every form factor
of the present nonzero-mass amplitude. This cancellation is conditional
on the endpoint domain and does not assert IR finiteness after the
RGZ masses are also taken to zero. Known real-mass results provide
checks of the supported integral limits through Package-X and
FeynHelpers \cite{Patel2015,Shtabovenko2017}.

\subsection{Infrared structure of the final scalar catalogue}
The IR poles of intermediate tensor coefficients should be
distinguished from those of the final scalar catalogue. The finite
triangle and box functions supplied here have the patterns in
Table~\ref{tab:scalarIR}. A single massless line does not give a
soft divergence at generic external kinematics. When two lines are
massless, their chord separation in this catalogue is a nonzero
invariant, rather than the zero invariant that would support a
collinear face.

\begin{table}[b]
\caption{Triangle and box families in the final catalogue. In each
row $r,r'$ are nonzero pole masses squared on the declared common
sheet, and $v\ne0$. IR regularity refers to this domain, not its
zero-mass or exceptional-kinematics boundaries.}
\label{tab:scalarIR}
\begin{ruledtabular}
\begin{tabular}{ll}
Family & Massless-line separation\\
\hline
$C_0(0,t,0;0,r,r')$ & Only one massless line\\
$C_0(v,0,0;0,0,r)$ & $v=s,u,t$\\
$D_0(v,0,t,0,0,0;0,0,r,r')$ & $v=s,u$\\
\end{tabular}
\end{ruledtabular}
\end{table}

More explicitly, near an isolated massless vertex of the Feynman
simplex let the remaining parameters be $\lambda y_i$. The scalar
measure behaves as $\lambda^{n-2}\dd\lambda$, and the denominator
as $\mathcal F^{D/2-n}$. If $\mathcal F=\lambda A(y)+O(\lambda^2)$
with nonvanishing slope on the relevant domain, their product is
$\lambda^{-\epsilon}\dd\lambda$, which is locally integrable.
Along the edge connecting two massless lines the nonzero chord
invariant prevents an identically vanishing denominator.
Continuation is understood away from Landau singularities.
Thus the final $C_0,D_0$ entries are both UV and IR finite in this
domain. All final bubbles have nonexceptional momentum and are IR
finite as well.

This observation does not recover a term discarded earlier as a
scaleless zero. Rather, it explains why the separately verified
intermediate IR cancellation is compatible with an IR-regular final
catalogue. The weighted ledger and ordered mass limits remain
necessary to establish the coefficients and finite transfers that
lead to that catalogue.

\subsection{Local counterterms and the UV pole}
\label{sec:renormalization}
The finite result is evaluated at $\alpha=0$ after
Eq.~\eqref{eq:gaugezero}. Set $a_s=g_s^2/(16\pi^2)$ and
$\Delta_{\UV}=1/\epsilon_{\UV}-\gamma_E+\ln4\pi$.
The parameter counterterms, with $N_f=2$, are
\begin{align}
 \frac{\delta g_s}{g_s}&=
 \left(-\frac{11N_c}{6}+\frac{N_f}{3}\right)a_s\Delta_{\UV},\notag\\
 \frac{\delta m^2}{m^2}&=
 \left(-\frac{35N_c}{12}+\frac{2N_f}{3}\right)a_s\Delta_{\UV},\notag\\
 \frac{\delta\mu^2}{\mu^2}&=-\frac{3N_c}{4}a_s\Delta_{\UV},\notag\\
 \frac{\delta m_G^4}{m_G^4}&=
 \left(-\frac{35N_c}{12}+\frac{2N_f}{3}\right)a_s\Delta_{\UV}.
 \label{eq:CT}
\end{align}
They follow from the Landau-gauge renormalization relations
\cite{Dudal2008,Capri2017}, together with
$\delta Z_A=(13N_c/6-2N_f/3)a_s\Delta_{\UV}$ and
Eq.~\eqref{eq:horizoncoordinate}. In particular,
\begin{align}
 & Z_{m^2}=Z_gZ_A^{-1/2},\quad
 Z_{\mu^2}=Z_gZ_A^{1/2},\notag\\
 &
 Z_{m_G^4}=Z_g^2(Z_{\gamma^2})^2,
 \quad Z_{\gamma^2}=Z_g^{-1/2}Z_A^{-1/4}.
\end{align}
The horizon counterterm is thus expressed in the independent
$m_G^4$ coordinate, not mistaken for a counterterm of $\gamma^4$.

The counterterm amplitude is constructed from the Born term:
\begin{equation}
 \mathcal M_{\rm ct}=\left(
 \delta g_s\partial_{g_s}+\delta m^2\partial_{m^2}
 +\delta\mu^2\partial_{\mu^2}
 +\delta m_G^4\partial_{m_G^4}\right)\mathcal M^{(0)}.
 \label{eq:CTamplitude}
\end{equation}
The external residues have already been included in
Eq.~\eqref{eq:assembly}; four additional quark field factors would
double count them. The coefficients in Eq.~\eqref{eq:CT} are not
fitted separately to scattering integrals.

An explicit check can be written compactly in the original color-flow
basis. Put $x=m^2$, $y=\mu^2$, $z=m_G^4$, and
$\Delta_t=(t-x)(t-y)+z$. The coefficient of $C_dS_1$ in the
unrenormalized one-loop UV pole is
\begin{align}
 U_d={}&\frac{g_s^4}{384\pi^2\Delta_t^2\epsilon_{\UV}}
 \bigg[\left(44-\frac{8N_f}{N_c}\right)(t-y)\Delta_t
 \notag\\[-2pt]
 &+\left(35-\frac{8N_f}{N_c}\right)
 \{x(t-y)^2-z(t-y)\}-9yz\bigg].
 \label{eq:UVexplicit}
\end{align}
For this calculation $N_f=2$. The $C_xS_1$ coefficient is $-N_cU_d$;
the other six color-flow/spin coefficients have vanishing UV poles.
These poles cancel Eq.~\eqref{eq:CTamplitude}. In the singlet/octet
basis the UV pole resides entirely in the octet $S_1$ form factor,
as required by the color structure of the tree counterterm.

\subsection{Finite terms and the measure}
A coefficient multiplying a singular integral must be expanded in
$D$ before its finite part is formed. If
\begin{align}
 c(4-2\epsilon)&=c_0+c_1\epsilon+c_2\epsilon^2+\cdots,\notag\\
 I&=\frac{I_{-2}}{\epsilon^2}+\frac{I_{-1}}{\epsilon}
 +I_0+\cdots,
\end{align}
then $[cI]_0=c_0I_0+c_1I_{-1}+c_2I_{-2}$.
The same rule is used when transforming tensor functions to the
scalar representation. The finite transformation includes
$-2c'(4)I_{-1}+2c''(4)I_{-2}$ and is not obtained by a
$D=4$ substitution into a generic-$D$ reduction identity.

The intermediate tensor integrals carry the fixed factor
$\ii\pi^2/(2\pi)^4$. Restoring the dimensional measure adds
$(2\pi)^{2\epsilon}$, whose linear term is
$2\epsilon\ln(2\pi)$. In the native PV convention this converts the
finite subtraction constant $-\gamma_E-\ln\pi$ into
$-\gamma_E+\ln4\pi$. This is a scheme conversion proportional to
the pole residue, not a second overall $1/(16\pi^2)$.
After the local subtraction and the separate IR cancellation,
Eq.~\eqref{eq:final} is finite in its stated domain.

\subsection{Scale dependence of the complete finite result}
\label{sec:scalecheck}
A check directly on the electronic finite coefficients is possible
without evaluating a complex-mass triangle or box. Let
$L_R=\ln\mu_R^2$, with parameters and kinematics held fixed.
On the common branch, the six explicit logarithms in the catalogue
have derivative $-1$, the nine remaining finite bubbles have
derivative $+1$, and the finite triangles and boxes have derivative
zero. The last statement uses the IR regularity just established,
not only their UV power counting. Therefore every row satisfies
\begin{equation}
 \partial_{L_R}F_{Ar}^{(1),\rm fin}
 =P_{Ar}\left[-\sum_{j=1}^{6}c_{Ar,j}
                    +\sum_{j=7}^{15}c_{Ar,j}\right].
 \label{eq:finiteScaleRow}
\end{equation}

To compare with the counterterms, write
$x=m^2$, $y=\mu^2$, $z=m_G^4$ and define
\begin{align}
 c_g&=-\frac{11N_c}{6}+\frac{N_f}{3},&
 c_x=c_z&=-\frac{35N_c}{12}+\frac{2N_f}{3},
 \notag\\
 c_y&=-\frac{3N_c}{4},&N_f&=2 .
\end{align}
In the convention of Eq.~\eqref{eq:CT}, the one-loop running with
respect to $L_R$ is
$\partial_{L_R}g_s=c_ga_sg_s$ and
$\partial_{L_R}(x,y,z)=a_s(c_xx,c_yy,c_zz)$.
Define the Born running operator
\begin{equation}
 \mathscr R_1=a_s\left(c_gg_s\partial_{g_s}
              +c_xx\partial_x+c_yy\partial_y+c_zz\partial_z\right).
\end{equation}
The eight coefficient rows give the exact identity
\begin{equation}
 \partial_{L_R}\mathcal M_{\rm fin}^{(1)}
 =-\mathscr R_1\mathcal M^{(0)} .
 \label{eq:RGclosure}
\end{equation}
Only the octet $S_1$ row has a nonzero scale derivative. This
identity was checked by rational algebra on all eight exported rows,
using $x=r_-+r_+-y$ and $z=r_-r_+-xy$. No numerical evaluation of a
loop integral or fit to a counterterm was used. Although the
coefficient identity holds formally in the displayed $N_c$, it
does not replace an independent arbitrary-group graph audit.

\begin{figure}[t]
\centering
\includegraphics[width=\columnwidth]{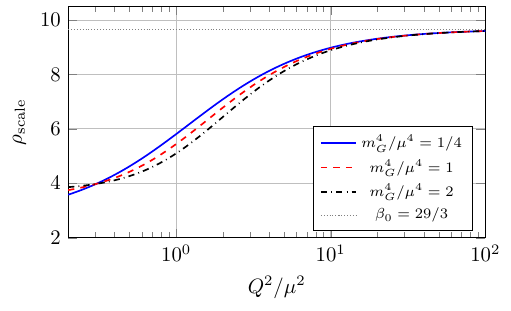}
\caption{Exact one-loop scale response from
Eq.~\eqref{eq:finiteScaleRow}, for $N_c=3$, $N_f=2$ and
$m^2/\mu^2=1$. Unlike the Born ratio, this probes the finite
one-loop coefficient table. It does not evaluate the scale-independent
parts of the multiscale functions and is not an inclusive observable.}
\label{fig:scale}
\end{figure}

The mass dependence of this one-loop scale response is compact.
For $B=-g_s^2 f(t)$, write
$\partial_{L_R}F_{81}^{(1),\rm fin}=a_s B\,\rho_{\rm scale}$.
Then
\begin{align}
 \rho_{\rm scale}={}&\beta_0
 -c_x\frac{x(t-y)}{\Delta_t}
 +c_y\frac{yz}{(t-y)\Delta_t}
 +c_z\frac{z}{\Delta_t},\notag\\
 \beta_0={}&\frac{11N_c-2N_f}{3}.
 \label{eq:rhoScale}
\end{align}
At spacelike exchange its asymptotic behavior is
$\rho_{\rm scale}=\beta_0+c_xm^2/Q^2+O(Q^{-4})$.
Figure~\ref{fig:scale} shows this quantity at fixed masses. It is a
response of the complete one-loop coefficient to the renormalization
scale, not a numerical value of the full finite amplitude.
Equation~\eqref{eq:RGclosure} cancels this response against the
running Born term through one loop.

\section{Mass dependence and interpretation}
\label{sec:massdependence}
The electronic form factors retain the full dependence on
$m^2,\mu^2,m_G^4$ through rational pole coordinates and finite scalar
integrals. They can be compared at two nonzero mass choices with the
same scale and analytic prescription. Equations~\eqref{eq:representativeC}
and \eqref{eq:representativeCmixed} already display a dependence
which cannot be obtained from a universal rescaling of the tree
propagator: different scalar integrals carry different functions of
the pole separation and residues.

\subsection{A spectral form of a one-loop triangle sector}
The three coefficients multiplying
$\bar C_0(0,t,0;0,r_\sigma,r_\tau)$ in the singlet $S_1$ row
can be combined before expanding pole residues. Define
$K=(s^2+t^2)/[s(s+t)]$ and
$C(x,y)=\bar C_0(0,t,0;0,x,y)$. The contribution inside $P_{11}$ is
\begin{equation}
 \mathcal T_{11}=-\frac K2
 \sum_{\sigma,\tau=\pm}a_\sigma a_\tau
 (r_\sigma+r_\tau-t)C(r_\sigma,r_\tau).
 \label{eq:spectralTriangle}
\end{equation}
Expanding this equation reproduces exactly the three printed
coefficients in the electronic table. The $a_\sigma$ are the
residues of Eq.~\eqref{eq:poles}, not extra adjustable weights.
This form displays how two full internal gluon lines generate the
mass dependence of an actual one-loop sector.

It also separates apparent spectral singularities from singularities
of the integral itself. If $r_\pm\to r$ on a nonsingular common sheet,
the single-line functional satisfies
\begin{equation}
 \sum_\sigma a_\sigma H(r_\sigma)
 \longrightarrow H(r)+(r-\mu^2)H'(r).
 \label{eq:spectralLimit}
\end{equation}
For the two-line function
$G(x,y)=(x+y-t)C(x,y)$ this gives
\begin{align}
 \mathcal T_{11}\longrightarrow-\frac K2\big[
 &G(r,r)+(r-\mu^2)(\partial_x+\partial_y)G(r,r)\notag\\
 &+(r-\mu^2)^2\partial_x\partial_yG(r,r)\big].
 \label{eq:triangleCoalescence}
\end{align}
The derivatives act on $G$ before setting $x=y=r$.
This follows from the double-pole limit
$f(v)\to1/(v-r)+(r-\mu^2)/(v-r)^2$ and has been checked by a
two-variable Taylor expansion. It is a conditional limit of this
triangle sector; it does not extend the certified domain of the
entire amplitude to coalescent poles.

\subsection{Projection onto a virtual interference quantity}
After pole cancellation and all finite dimensional transfers, the
four-dimensional spin sums provide a useful route from form factors
to a scalar comparison. Put
$K_r=\tfrac14\sum_{\rm spins}S_1^*S_r$. In the conventions of
Eq.~\eqref{eq:spin}, direct traces give
\begin{align}
 K_1&=2(s^2+u^2),&
 K_2&=-su(s-u),\notag\\
 K_3&=8(4s^2+u^2),&
 K_4&=4s(2s^2+u^2).
 \label{eq:spinInterference}
\end{align}
These traces were also checked with explicit four-by-four Dirac
matrices at three nonexceptional scattering angles. The projection
is not performed on the divergent $D$-dimensional amplitude.

Color orthogonality implies
\begin{equation}
 \sum_{\rm colors}C_1^*C_8=0,\qquad
 \sum_{\rm colors}|C_8|^2=\frac{N_c^2-1}{4}.
\end{equation}
Consequently, for $B=-g_s^2f(t)$ the normalized virtual interference
is
\begin{equation}
 \delta_{\rm virt}(s,t)=
 \frac{2\,\Re\!\left[
 B^*\sum_{r=1}^4K_rF_{8r}^{(1),\rm fin}\right]}
 {|B|^2K_1}.
 \label{eq:virtualInterference}
\end{equation}
The singlet coefficients are nonzero parts of the amplitude but
do not interfere with the Born term in this color-summed quantity.
In particular, the explicit singlet logarithms of
Eq.~\eqref{eq:hardlogs} alone are not an NLO interference prediction.
At fixed parameters,
$\partial_{L_R}\delta_{\rm virt}
=2a_s\Re\rho_{\rm scale}$.
Equation~\eqref{eq:virtualInterference} is an evaluable virtual
projection, not a result for an inclusive cross section.
Section~\ref{sec:numerics} supplies a common-sheet evaluator and
numerical values of its scale-independent part.

\subsection{Tree-level reference and massless boundary}

A useful normalization check is available analytically at Born
level. For spacelike exchange $t=-Q^2$, define the ratio to the
massless exchange kernel,
\begin{equation}
 \mathcal R(Q^2)=
 \frac{Q^2(Q^2+\mu^2)}
 {(Q^2+m^2)(Q^2+\mu^2)+m_G^4}.
 \label{eq:BornRatio}
\end{equation}
It approaches unity at large $Q^2$, with
\begin{align}
 \mathcal R(Q^2)={}&1-\frac{m^2}{Q^2}
 +\frac{m^4-m_G^4}{Q^4}\notag\\
 &+\frac{-m^6+(2m^2+\mu^2)m_G^4}{Q^6}
 +O(Q^{-8}).
 \label{eq:highenergyBorn}
\end{align}
In particular the auxiliary mass first enters this expansion through
the horizon contribution at order $Q^{-6}$. This is a statement
about the Born exchange factor, not a high-energy expansion of the
complete one-loop correction.

\begin{figure}[t]
\centering
\includegraphics[width=\columnwidth]{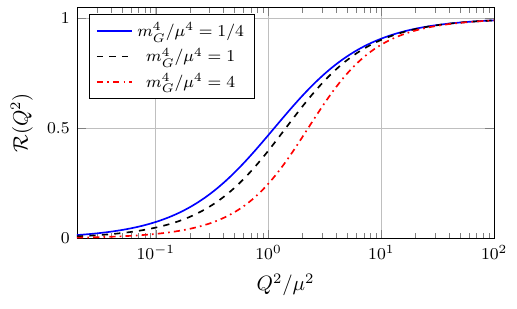}
\caption{Born exchange ratio, Eq.~\eqref{eq:BornRatio}, for
$m^2/\mu^2=1$ and three horizon mass choices. These illustrative
curves use the exact tree-level formula and contain no one-loop
numerical integral evaluation.}
\label{fig:born}
\end{figure}

The formal spin- and color-averaged Born quantity is
\begin{equation}
 \br{|\mathcal M^{(0)}|^2}=
 \frac{N_c^2-1}{2N_c^2}g_s^4(s^2+u^2)|f(t)|^2.
 \label{eq:bornsquared}
\end{equation}
Consequently its ratio to the corresponding massless expression
is $|\mathcal R(Q^2)|^2$. Figure~\ref{fig:born} illustrates this
limited diagnostic. It is not a numerical evaluation of the one-loop
form factors or a fit of the RGZ masses.

The simultaneous QCD massless limit is not a point within the
nonzero-mass IR domain used above. New soft and collinear singularities
can emerge, while individual spectral coefficients become singular
when poles coalesce. A QCD comparison must therefore be taken on
complete integral combinations with a common IR regulator; it cannot
be made by replacing every $r_\pm$ and $\mu^2$ by zero in the
finite coefficient table. Nor does the finite virtual amplitude by
itself define an inclusive next-to-leading-order cross section.
Real emission, an IR-safe observable, and the corresponding
factorization prescription must be specified for that purpose.

\section{Physical-region evaluation and the QCD boundary}
\label{sec:numerics}
\subsection{A common-sheet numerical benchmark}
We use the corrected coefficients and the same continued scalar
functions in the supplied amplitude,
rather than only its scale derivative. Introduce a reference mass $m_0$
and the illustrative path
\begin{equation}
 \begin{gathered}m^2=\mu^2=\lambda^2m_0^2,\quad m_G^4=\lambda^4m_0^4/4,\\
 r_\pm=\lambda^2m_0^2(1\pm i/2).\end{gathered}
 \label{eq:num-path}
\end{equation}
The two spectral weights are $1/2$. We set $N_c=3$, $N_f=2$,
$\mu_R^2=m_0^2$, and $t=u=-s/2$. These choices illustrate the
mass dependence; they are not a fit to hadronic data.

The evaluation retains the Feynman sheet, in particular
$\log(-s-i0)=\log s-i\pi$. Conjugating pole masses alone must not
conjugate this kinematic prescription. For example, two useful
representations, with squared masses $a,b$, are
\begin{align}
 &C_0(0,t,0;0,a,b)\notag\\
 &\quad=\int_0^1\!\dd x\,
 \frac{\log[1-tx(1-x)/(ax+b(1-x))]}{tx(1-x)},\notag\\
 &v C_0(v,0,0;0,0,a)=\operatorname{Li}_2(-v/a)\notag\\
 &\quad-(\log a-\log(-v-i0))\log(1+v/a).
 \label{eq:num-triangles}
\end{align}
The dimensionful logarithms in the second line appear only as a
ratio. Boxes are evaluated by a Cheng--Wu parameterization and an
analytic radial integration, as detailed in the Supplemental Material.
For the phases in Eq.~\eqref{eq:num-path}, the radial contour can be
rotated through $-\pi/4$ while retaining the negative imaginary part
of the denominator. This specifies continuation from the Euclidean
region for these benchmarks, rather than a prescription for arbitrary
complex masses.

\begin{figure}[t]
\centering
\includegraphics[width=\columnwidth]{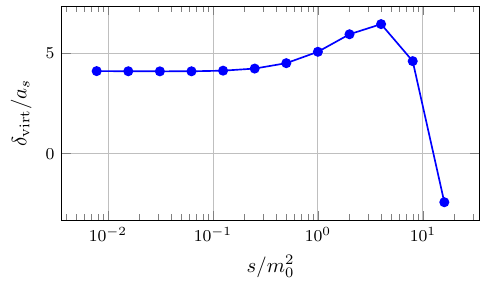}
\caption{Finite virtual correction on Eq.~\eqref{eq:num-path} at
$\lambda=1$, $t=u=-s/2$, and $\mu_R^2=m_0^2$. All finite integral
families are evaluated on the same continued sheet.}
\label{fig:virtual-numeric}
\end{figure}

Figure~\ref{fig:virtual-numeric} shows the complete finite virtual
interference, normalized by $a_s=g_s^2/(16\pi^2)$, at $\lambda=1$.
Unlike a propagator ratio, it includes all eight finite form factors.
Its scale-independent part is substantial: at $s/m_0^2=4$,
$\delta_{\rm virt}/a_s=6.42863942$. The curve is only a guide through
twelve computed points. This finite virtual quantity is not an inclusive
NLO cross section.

Table~\ref{tab:numeric-ff} supplies a reproducible amplitude benchmark.
Define $\widehat F_{cr}=16\pi^2m_0^{2d_r}F_{cr}/g_s^4$, where
$d_{1,3}=1$ and $d_{2,4}=2$. Here the color labels $1,8$ denote the
singlet and octet basis of the main text. An independent integration
order and rotated contour for all six boxes changes the eight
form factors by at most $1.7\times10^{-13}$ in these units and
$\delta_{\rm virt}/a_s$ by $2.6\times10^{-11}$ at this point.
The color correction changes no box coefficient, so this scalar
integration-order comparison remains applicable after the coefficient
update. This is an integration consistency test, not a new graph-generation
or gauge-independence proof. We quote fewer digits than the internal
working precision.
\begin{table}[t]
\caption{Finite coefficients at $s=4m_0^2$, $t=-2m_0^2$, $\lambda=1$.}
\label{tab:numeric-ff}
\begin{ruledtabular}\begin{tabular}{ccrr}
Color & Spin & $\Re\widehat F$ & $\Im\widehat F$\\
1&1&0.248214941&$-0.295449508$\\
8&1&2.185465652&0.443174262\\
1&2&$-0.012685572$&$-0.021638300$\\
8&2&0.064984183&0.032457450\\
1&3&$-0.052743801$&$-0.028997427$\\
8&3&$-0.112336754$&0.043496141\\
1&4&0.014404470&$-0.000025709$\\
8&4&0.021306615&0.000038564\\
\end{tabular}\end{ruledtabular}
\end{table}

\subsection{Ordered massless limit and a finite-hard comparison}
The finite four-dimensional answer cannot be evaluated at zero masses
to obtain dimensionally regulated QCD. A simple example shows the
noncommuting limits. In the $\MS$ normalization, a regulated triangle is
\begin{align}
 C(v,a;\epsilon)={}&-\frac{e^{\gamma_E\epsilon}\Gamma(1+\epsilon)
 (\mu_R^2)^\epsilon}{\epsilon}
 \int_0^1\!\dd x\,(1-x)^{-\epsilon}\notag\\[-2mm]
 &\hspace{3mm}\times
 \frac{(-vx-i0)^{-\epsilon}-a^{-\epsilon}}{a+vx}.
 \label{eq:ordered-triangle}
\end{align}
At fixed $\Re\epsilon<0$, its $a\to0$ limit is
\begin{align}
 &\frac{e^{\gamma_E\epsilon}\Gamma(1+\epsilon)\Gamma(1-\epsilon)^2}
 {v\epsilon^2\Gamma(1-2\epsilon)}
 \left(\frac{\mu_R^2}{-v-i0}\right)^\epsilon\notag\\
 &\quad=\frac1v\left[\frac1{\epsilon^2}-\frac{L_v}{\epsilon}
 +\frac{L_v^2}{2}-\frac{\pi^2}{12}\right]+O(\epsilon),
 \label{eq:massless-triangle}
\end{align}
where $L_v=\log[(-v-i0)/\mu_R^2]$. Independent parameter
quadrature verifies this ordered primitive. It explains why the
finite-mass amplitude and the usual QCD finite hard amplitude require
an IR conversion before comparison.

For physical massless external helicities the spin structures obey
\begin{equation}
 \frac{(\mathcal S_1,\mathcal S_2,\mathcal S_3,\mathcal S_4)}
 {\mathcal S_1}=
 \begin{cases}(1,-u/2,16,4s),&LL,\\
 (1,s/2,4,2s),&LR.
 \end{cases}
 \label{eq:chiral-ratios}
\end{equation}
Parity supplies $RR,RL$. Explicit Dirac matrices verify these
relations at three nonexceptional scattering angles. They are used
only after the finite $D$-dimensional answer has been assembled;
no evanescent structure is discarded before pole subtraction.

We compare all four physical color--helicity coefficients to the
massless QCD hard coefficients of Ref.~\cite{KelleySchwartz2011}.
At leading power on Eq.~\eqref{eq:num-path}, define
\begin{align}
 \ell&=\log\lambda^2+\tfrac12\log(5/4),&
 \ell_2&=\ell^2-\arctan(1/2)^2,\notag\\
 \mathcal D&=-\tfrac13L_s+\tfrac16L_t-\tfrac76L_u.
\end{align}
The one-loop soft/collinear conversion, relative to the massless
Born octet coefficient, is
\begin{align}
 Z_8={}&-2C_F\ell_2+(-4\mathcal D-6C_F)\ell
       +2C_F\left(\tfrac92-\tfrac{5\pi^2}{6}\right),\notag\\
 Z_1={}&-\frac{2C_F}{C_A}(L_s-L_u)\ell .
 \label{eq:ir-conversion}
\end{align}
Here $C_F=4/3$, $C_A=3$. The dipole factors are those of the
four-quark color basis; their sum is $-C_F$. The two logarithmic
moments are spectral averages of the single massive-vector regulator.
The constant follows from the massive-vector quark form factor
\cite{Feucht2004} relative to the dimensionally regulated massless
hard form factor. It is fixed independently of the four-quark fit.
This conversion assumes the leading soft/collinear factorization
of minimally coupled quarks and the perturbative zero-mass branch;
it is not a proof of factorization in every RGZ parameter domain.

Let $V_{ch}$ be the computed finite helicity coefficient divided by
$a_s(-g_s^2/t)$. Subtract $Z_8$ from both octet helicities and
$Z_1$ from both singlet helicities. At $s=4m_0^2$, $t=u=-2m_0^2$,
we evaluate nine masses from $\lambda=1$ to $1/256$.
The four smallest masses are fitted with a free intercept,
\begin{equation}
 H_{ch}(\lambda)=H_{ch}(0)+\lambda^2
 [c_0+c_1\log\lambda^2+c_2\log^2\lambda^2].
 \label{eq:qcd-extrapolation}
\end{equation}
The intercept is not fixed to QCD. Table~\ref{tab:qcd-hard} compares
it with the independently specified reference. The maximum complex
absolute difference is $3.6\times10^{-6}$; using five masses changes
the intercept by at most $9.0\times10^{-6}$. We therefore interpret
this as agreement at about $2\times10^{-5}$ absolute accuracy,
allowing for fit variation, rather than as an exact symbolic identity.
The unextrapolated $1/256$ coefficients still differ by up to
$2.8\times10^{-3}$. Machine precision in the final coefficient
assembly and quadrature errors also preclude an arbitrary-precision
claim based solely on the integration working precision.
\begin{table}[t]
\caption{QCD hard boundary at $s=4m_0^2$, $t=u=-2m_0^2$,
$\mu_R^2=m_0^2$, $N_f=2$. Entries are dimensionless coefficients
relative to $a_s$ times the massless Born octet amplitude.}
\label{tab:qcd-hard}
\begin{ruledtabular}\begin{tabular}{crr}
 & QCD reference & Free-intercept limit\\
$8,LL$&$27.060121+2.903448i$&$27.060122+2.903448i$\\
$8,LR$&$35.714379+0.642671i$&$35.714383+0.642671i$\\
$1,LL$&$-1.217865-1.935632i$&$-1.217865-1.935632i$\\
$1,LR$&$0.414833-0.428447i$&$0.414833-0.428447i$\\
\end{tabular}\end{ruledtabular}
\end{table}
This tests the complete physical helicity amplitude at one
kinematic point along one common scaling path. It supplements the
symbolic gauge check and finite-mass coefficient identities; it
does not establish analytic QCD matching at general kinematics or
for independently vanishing masses. The electronic records retain
all scan points and both fit windows.

\subsection{QCD checks away from the symmetric angle}
\label{sec:asymmetric-qcd}
An equal-$t,u$ point can conceal interchange errors or kinematic
identities absent at generic scattering angle. We therefore repeat
the complete four-component comparison at $s=4m_0^2$ with
$\zeta=-t/s=1/3$ and $2/3$, retaining $u=-s-t$ separately.
For each angle all finite functions are evaluated at
$\lambda=1/16,1/32,1/64,1/128,1/256$ on the same mass path and
continued sheet. The subtraction in Eq.~\eqref{eq:ir-conversion}
uses the corresponding distinct $L_t,L_u$, and the reference is the
angle-dependent hard amplitude of Ref.~\cite{KelleySchwartz2011}.
All four coefficients, including both their real and imaginary
parts, are compared. No intercept is fixed to its QCD value.

\begin{table}[t]
\caption{QCD-boundary comparison at three scattering angles.
$E_4$ is the maximum complex absolute difference between the
four-point free intercept and the QCD reference, over all four
components. $W$ is the maximum change when five mass points are
used. $E_{\lambda}$ is the unextrapolated difference at $\lambda=1/256$.}
\label{tab:qcd-angles}
\begin{ruledtabular}\begin{tabular}{cccc}
$\zeta$&$E_4$&$W$&$E_{\lambda}$\\
$1/3$&$6.10\times10^{-6}$&$1.28\times10^{-5}$&$4.50\times10^{-3}$\\
$1/2$&$3.57\times10^{-6}$&$9.00\times10^{-6}$&$2.79\times10^{-3}$\\
$2/3$&$2.81\times10^{-6}$&$7.83\times10^{-6}$&$1.86\times10^{-3}$\\
\end{tabular}\end{ruledtabular}
\end{table}
Table~\ref{tab:qcd-angles} supports numerical agreement at an
absolute level of about $3\times10^{-5}$, allowing for the observed
fit-window variation. The five-point fits differ from the reference
by at most $1.89\times10^{-5}$. This is an empirical accuracy
assessment, not a rigorous extrapolation-error bound. The visibly
nonzero finite-mass residuals remain in the electronic data; they
are not truncated to zero. The twelve complex limiting coefficients
and every mass point are supplied in
\path{data/qcd-angles-comparison.csv} and the corresponding WXF
packet. These checks remove the restriction to the symmetric angle
from the numerical comparison, while retaining a single mass path
and three selected kinematic points. They do not prove analytic
matching for arbitrary invariants or independent massless paths.

\subsection{Infrared changes relative to massless QCD}
\label{sec:infrared-physics}
The massless boundary check alone does not describe the region where
RGZ masses compete with the transferred momentum. We therefore extend
the finite-amplitude scan down to $s/m_0^2=1/128$, holding
$m^2=\mu^2=m_0^2$ and $m_G^4=m_0^4/4$ fixed. The distinction
between a finite-mass amplitude and an IR-subtracted massless hard
coefficient is essential here: the leading-power small-mass conversion
in Eq.~\eqref{eq:ir-conversion} is not extrapolated into this region.

At tree level, spacelike exchange becomes a local current interaction:
\begin{align}
 B(-Q^2)&=\frac{g_s^2\mu^2}{m^2\mu^2+m_G^4}+O(Q^2),\notag\\
 \mathcal R(Q^2)&=\frac{\mu^2 Q^2}{m^2\mu^2+m_G^4}+O(Q^4).
 \label{eq:ir-contact}
\end{align}
For fixed nonexceptional angle the spinor currents contribute one
power of $s$; thus the Born amplitude vanishes linearly with $s$
instead of retaining the massless Coulomb exchange behavior. On our
path, with $\rho=s/m_0^2$ and $Q^2=s/2$,
\begin{equation}
 |\mathcal R|^2=\frac{4\rho^2}{25}+O(\rho^3).
 \label{eq:ir-born-scaling}
\end{equation}
At $\rho=1/128$ the exact ratio is $9.71995\times10^{-6}$.
This ratio compares spin/color averaged Born quantities; it is not
an inclusive QCD NLO ratio.

A loop-level comparison can avoid a helicity-blind IR conversion
entirely. Denote the coefficient of $C_c\mathcal S_1$ in a physical
helicity amplitude by $A_{c,h}$. Define
\begin{align}
 \mathcal D_c&=\frac{A^{(1)}_{c,LR}-A^{(1)}_{c,LL}}
 {a_s B_{\rm QCD}},& B_{\rm QCD}&=-g_s^2/t,\notag\\
 \mathcal D_c&=\frac{(-t/2)F_{c2}-12F_{c3}-2sF_{c4}}
 {a_s B_{\rm QCD}}.
 \label{eq:ir-chiral-difference}
\end{align}
The identical Born coefficients cancel in this difference. The
one-loop soft color operator and collinear factors act on the Born
amplitude without resolving quark helicity. Consequently the
massless IR poles and any common helicity-blind finite IR conversion
also cancel between $LL$ and $LR$ in each color component
\cite{KelleySchwartz2011}. In the RGZ result the exact scale slope
is confined to $F_{81}$, which drops out of
Eq.~\eqref{eq:ir-chiral-difference}. These differences therefore
have zero explicit one-loop scale slope, verified on the coefficient
table. They are finite amplitude diagnostics, not cross sections.

At $t=u=-s/2$ the QCD reference is scale independent and reads
\begin{align}
 \mathcal D_8^{\rm QCD}={}&\log^22-\tfrac23\log2+\tfrac78\pi^2
 +i\pi(\tfrac23-2\log2),\notag\\
 \mathcal D_1^{\rm QCD}={}&-\tfrac23\log^22+\tfrac49\log2+\tfrac16\pi^2
 +i\pi(\tfrac43\log2-\tfrac49).
 \label{eq:ir-qcd-target}
\end{align}
Numerically these are $8.654259-2.260777i$ and
$1.632697+1.507185i$. Unlike the individual hard coefficients,
these combinations need no extrapolated small-mass IR subtraction
to be compared with the finite RGZ result.

\begin{figure}[t]
\centering
\includegraphics[width=\columnwidth]{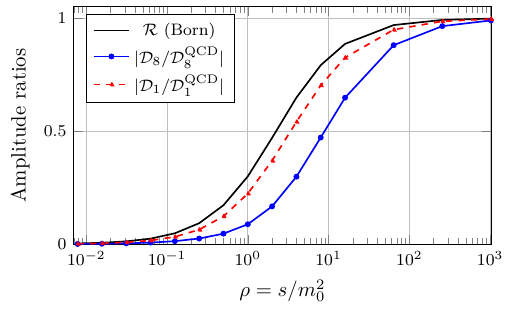}
\caption{Infrared suppression and approach to QCD on the common
mass path at $t=u=-s/2$. The loop curves use the helicity differences
in Eq.~\eqref{eq:ir-chiral-difference}; no asymptotic IR conversion
is applied at small $\rho$. Curves join computed points.}
\label{fig:ir-shapes}
\end{figure}

Figure~\ref{fig:ir-shapes} separates the tree exchange effect from
new loop structures. At $\rho=1/128$ we obtain
\begin{align}
 \mathcal D_8&=0.00830599-0.0000271114i,\notag\\
 \mathcal D_1&=0.00450861+0.0000180743i.
 \label{eq:ir-shape-point}
\end{align}
Their magnitudes relative to QCD are $9.2860\times10^{-4}$ and
$2.0291\times10^{-3}$, whereas $\mathcal R=3.11768\times10^{-3}$.
In particular,
\begin{align}
 \frac{\mathcal D_8}{\mathcal R\mathcal D_8^{\rm QCD}}
 &=0.28842+0.07434i,\notag\\
 \frac{\mathcal D_1}{\mathcal R\mathcal D_1^{\rm QCD}}
 &=0.47999-0.43954i.
 \label{eq:ir-not-rescaling}
\end{align}
A universal replacement of the Born propagator cannot produce
these different complex ratios. Across the sampled deep-IR points
the real parts decrease approximately linearly with $\rho$ and
the imaginary parts more rapidly. We do not infer exact all-order
powers or an IR factorization theorem from this finite scan.
At large $\rho$ both loop ratios approach unity, providing a
QCD comparison free of the common conversion used for the four
individual helicity coefficients.

The relative virtual correction also remains moderate in the
sampled IR region at the fixed subtraction scale $\mu_R^2=m_0^2$:
$\delta_{\rm virt}/a_s=4.09797$ at $\rho=1/128$ and $4.08841$
at $\rho=1/16$. With the illustrative choice $g_s=1$, the former
corresponds to a $2.60\%$ correction relative to the suppressed
RGZ Born quantity. Relative to the QCD Born quantity, the
RGZ Born plus virtual interference is $9.97219\times10^{-6}$.
This is a fixed-coupling coefficient study, not a prediction for
the running coupling in deep-infrared QCD.

Changing the subtraction scale to $\mu_R^2=s$ while holding the
renormalized parameters fixed shifts
\begin{equation}
 \frac{\delta_{\rm virt}(s)}{a_s}
 =\frac{\delta_{\rm virt}(m_0^2)}{a_s}
 +2\rho_{\rm scale}\log(s/m_0^2).
 \label{eq:ir-scale-shift}
\end{equation}
For example, at $\rho=1/128$ this gives $-22.2833$, rather than
$4.09797$. This change is explicit scale dependence, compensated
by parameter running at the appropriate order; it must not be
interpreted as a new mass effect. The helicity differences above
avoid this particular ambiguity.

\begin{table}[t]
\caption{IR variations at $s=m_0^2/16$, $\mu_R^2=m_0^2$,
$m^2=\mu^2=m_0^2$. Here $z=-t/s$, $h=m_G^4/m_0^4$.
All entries are evaluated, not interpolated.}
\label{tab:ir-variations}
\begin{ruledtabular}\begin{tabular}{ccrr}
$z$&$h$&$\mathcal R$&$\delta_{\rm virt}/a_s$\\
$1/4$&$1/4$&0.0123833&4.756110\\
$1/2$&$1/4$&0.0245353&4.088405\\
$3/4$&$1/4$&0.0364593&3.501947\\
$1/2$&$1/16$&0.0286210&4.415485\\
$1/2$&$1$&0.0156176&3.080947\\
\end{tabular}\end{ruledtabular}
\end{table}
Table~\ref{tab:ir-variations} checks that the result is sensitive
to angle and horizon strength, not only to a common unit change.
All displayed phases admit the stated rotated-contour continuation.
An independent integration order for all six boxes at
$s=m_0^2/32$ changes $\delta_{\rm virt}/a_s$ by
$2.7\times10^{-11}$. Independently evaluated points with the same
$s/(\lambda^2m_0^2)$ agree after the exact scale shift to within
$9\times10^{-12}$. These tests control numerical consistency;
they do not replace general parameter scans or phenomenological
constraints on the RGZ masses. In particular, the perturbative QCD
reference at arbitrarily low momentum is a formal coefficient
comparison, not a reliable hadronic cross section.

\section{Physical interpretation}
\label{sec:interpretation}
\subsection{A low-momentum interaction scale, rather than a single gluon mass}
The infrared modification can be organized without interpreting either
complex pole as an observable particle mass. For positive benchmark
parameters, set $x=m^2$, $y=\mu^2$, $z=m_G^4$ and $d_0=xy+z$.
The exchange coefficient has the expansion
\begin{align}
 B(-Q^2)&=\frac{g_s^2}{\Lambda_{\rm ex}^2}
 \left[1+c_{\partial}Q^2+O(Q^4)\right],\notag\\
 \Lambda_{\rm ex}^2&=x+\frac{z}{y},&
 c_{\partial}&=\frac{z-y^2}{y(xy+z)}.
 \label{eq:contact-matching}
\end{align}
The leading term matches the matrix element of the octet vector
current product $(\bar q\gamma_\mu T^a q)
(\bar q'\gamma^\mu T^a q')$, in our amplitude convention.
Its coefficient has mass dimension $-2$; the next term describes
momentum-dependent corrections. This is tree-level matching, not a
complete Wilsonian integration of the massless sectors of the theory.
Its Taylor regime requires momenta small compared with the nonzero
pole scales, not merely small compared with a subtraction scale.

At leading order in this expansion a low-momentum measurement of
the exchange strength determines only $\Lambda_{\rm ex}$.
It cannot separately determine $m^2$, $\mu^2$ and $m_G^4$.
The momentum slope and loop color--spin structures supply additional
independent information. Thus the parameters should not be inferred
from a single normalization fit or identified with a single effective
mass inserted into QCD diagrams.

Two further diagnostics follow exactly. The exchanged coefficient
obeys
\begin{equation}
 \frac{\partial B(-Q^2)}{\partial Q^2}
 =g_s^2\frac{z-(Q^2+y)^2}
 {[(Q^2+x)(Q^2+y)+z]^2}.
 \label{eq:exchange-slope}
\end{equation}
For $z\leq y^2$ it decreases with spacelike momentum; for $z>y^2$
it first increases and reaches a maximum at $Q^2=\sqrt z-y$.
This is a property of the Euclidean exchange coefficient, not a
physical resonance. The amplitude ratio to massless exchange reaches
$\mathcal R=1/2$ at
\begin{equation}
 Q_{1/2}^2=\frac{x-y+\sqrt{(x+y)^2+4z}}{2}.
 \label{eq:half-exchange}
\end{equation}
For the reference masses this is $\sqrt5\,m_0^2/2$.
Consequently the transition is controlled by a combination of masses,
and the different curves in Fig.~\ref{fig:ir-shapes} need not turn
at an identical momentum.

\subsection{Angular redistribution and the meaning of infrared suppression}
Let $\zeta=-t/s$ be fixed away from 0 and 1. Equations~\eqref{eq:bornsquared}
and \eqref{eq:contact-matching} give
\begin{equation}
 \overline{|\mathcal M^{(0)}|^2}
 =\frac{N_c^2-1}{2N_c^2}
 \frac{g_s^4s^2}{\Lambda_{\rm ex}^4}
 [1+(1-\zeta)^2]+O(s^3).
 \label{eq:ir-angular}
\end{equation}
For a formal massless two-body phase space,
$\dd\sigma/\dd\Omega=\overline{|\mathcal M|^2}/(64\pi^2s)$,
the Born contribution therefore scales as $s$ at fixed angle,
whereas massless QCD Born exchange scales as $1/s$.
Their ratio vanishes as $s^2$. This is the kinematic origin of
the strong suppression in the IR scan; it is not inferred from
an arbitrarily small coupling.

There is also an angular change. At fixed $s$, the massless
$1/t^2$ enhancement of the Born squared amplitude is absent when
$t\to0$ with the RGZ masses nonzero. The remaining leading
angular factor is $1+(1-\zeta)^2$. Thus the mass effect changes
both normalization and the forward concentration of tree exchange.
This statement concerns the Born term. Massless loop regions can
retain nonanalytic angular dependence, so the fixed-angle scan does
not establish a uniform forward limit for the full amplitude.
Nor should the formal two-body scaling be read as a hadronic or
inclusive NLO prediction.

\subsection{Spin information survives the suppressed exchange}
The quantities $\mathcal D_c$ are normalized to the QCD Born
coefficient. To expose the intrinsic RGZ correction, define instead
\begin{equation}
 \Xi_c=\frac{\mathcal D_c}{\mathcal R}
 =\frac{A^{(1)}_{c,LR}-A^{(1)}_{c,LL}}{a_s B}.
 \label{eq:intrinsic-spin}
\end{equation}
At $s/m_0^2=1/128$ the computed values are
\begin{align}
 \Xi_8&=2.66415-0.00869602i,\notag\\
 \Xi_1&=1.44614+0.00579735i.
 \label{eq:intrinsic-spin-point}
\end{align}
Although $\mathcal D_c$ is small, these coefficients are not.
The loops modify a weakened exchange rather than simply switching
off. The distinction between $LL$ and $LR$ is a chirality-conserving
spin correlation; it does not imply a generated quark mass or
spontaneous chiral-symmetry breaking. In a local low-energy
organization, parity permits separate coefficients for the
$LL+RR$ and $LR+RL$ current products. A vector-current Born
interaction gives them equal coefficients; loops can distinguish
them while preserving parity and chirality on each massless line.
The full one-loop amplitude also retains massless regions and
nonanalytic terms, so Eq.~\eqref{eq:intrinsic-spin-point} is a
finite-kinematic diagnostic, not an independently matched local
Wilson coefficient.

The singlet amplitude has a different role from the octet correction.
With the color basis of Eq.~\eqref{eq:color},
\begin{equation}
 \frac1{N_c^2}\sum_{\rm colors} C_1^*C_8=0,\qquad
 \frac1{N_c^2}\sum_{\rm colors}|C_8|^2
 =\frac{N_c^2-1}{4N_c^2}.
 \label{eq:color-interference-meaning}
\end{equation}
Since the Born amplitude is purely octet, a one-loop singlet does
not contribute to its color-averaged interference. Its direct square
starts at $g_s^8$, beyond the $g_s^6$ virtual interference studied
here. This does not make the singlet redundant: it is an independent
component of the amplitude and of subsequent color evolution
\cite{KelleySchwartz2011}. A complete result at order $g_s^8$
would also need other contributions, including two-loop interference;
we do not promote the singlet square alone to such a prediction.

\subsection{Phases, mass sensitivity, and perturbative scope}
The surviving imaginary parts show why finiteness is not equivalent
to analyticity at zero external energy. Massless quarks and other
massless sectors remain in the calculation. Timelike logarithms and
integrals retain their continued discontinuities even though the
gluon exchange is finite at $Q^2=0$. The sampled imaginary parts of
$\mathcal D_c$ decrease faster than their real parts, consistently
with additional low-energy phase-space suppression, but assigning
this behavior to particular physical cuts would require a separate
cut-diagram analysis. Complex conjugate gluon poles themselves are
not positive-norm asymptotic states or measured decay widths. The
present continuation does not by itself establish an optical theorem
for a colored-particle S matrix. BRST gauge independence and such a
unitarity statement are distinct questions~\cite{Capri2016}.

The horizon-mass variation helps separate mechanisms. At
$s=m_0^2/16$ and $-t/s=1/2$, increasing $h=m_G^4/m_0^4$ from
$1/16$ to $1$ changes $\mathcal R$ from $0.0286210$ to
$0.0156176$. The fixed-scale virtual coefficient changes from
$4.41533$ to $3.06655$. The former is directly understood from
$\Lambda_{\rm ex}^2=m_0^2(1+h)$; the latter additionally probes
vertices, boxes and mixed self-energies, and is not fixed by the
exchange strength. These are illustrative parameter variations,
not fitted evidence for one preferred horizon mass.

Finally, a small relative correction at $g_s=1$ tests only this
chosen perturbative organization. If a coefficient of size four
multiplies $a_s$, keeping that contribution below ten percent would
require roughly $a_s<0.025$; it does not follow that a physical
running coupling obeys that bound. Large logarithms, exceptional
angles or a changed hierarchy can spoil the hierarchy even when
individual integrals remain finite. The scale check and the
scale-insensitive combinations $\mathcal D_c$ are useful controls,
while a running-parameter analysis and real radiation are required
for further phenomenological applications.

\section{Conclusions}
We have constructed the distinct-flavor massless one-loop four-quark
amplitude of a local RGZ action with its constraint determinant and
two dynamical quark flavors. The computation includes the complete
mixed self-energy attachment, the full auxiliary-field non-Abelian
vertices, both box routings, and the external residues. Its symbolic
gauge difference vanishes after the regulator is removed in the
common dimensional prescription. UV/IR bookkeeping and local
counterterms then yield a finite representation at nonzero RGZ
masses, with the limiting order and analytic domain retained.

The corrected result changes fourteen coefficients in the octet
Born spin structure relative to the earlier table. The other seven
rows are identical, while the finite virtual interference must be
updated. The eight color--spin form factors are given by a
thirty-entry scalar integral inventory. All 144 coefficients of the
finite representation are supplied with an exact reconstruction
check. This makes the surviving gauge-independent mass dependence
accessible without expanding very large open-spinor expressions.
It also isolates the physical content of the calculation from
individual gauge-dependent graph rearrangements.

The finite coefficients also satisfy an exact scale-response identity:
all scale dependence is confined to the Born octet structure and is
cancelled by coupling and mass running. A compact two-line spectral
triangle formula exhibits the origin of the mass weights and its
conditional double-pole limit. The four-dimensional interference
projection yields a complete finite virtual numerical benchmark.
A common-sheet evaluation and a controlled small-mass scan test all
four physical color--helicity coefficients against QCD at three
scattering angles, after an independently specified IR conversion.
A lower-momentum scan further separates tree suppression from
color--helicity effects, using differences in which common IR and
explicit one-loop scale terms cancel.
The contact expansion identifies a combined low-momentum interaction
scale; the surviving loop spin differences and the singlet color
component carry information beyond this exchange normalization.

Further applications require additional ingredients according to the
observable of interest. An analytic QCD matching proof at general
kinematics, continuation beyond the benchmark mass-phase domain,
and inclusive radiation remain distinct tasks. The
present result supplies the virtual partonic input on its stated
branch and domain, without a claim about nonperturbative colored
asymptotic states or all-order functional identities.

\section*{Data availability}
The corrected 144 coefficients, exact new-minus-old differences,
updated numerical data, and executable checks accompany this revision.
The new result is bound to the color-corrected conditional certificate
and its input files by SHA-256 records. The companion implementation
is RGZWorkflow version 1.1.2 under GPL-3.0-or-later~\cite{WuRGZWorkflow112}; its exact source
release is archived at \href{https://doi.org/10.5281/zenodo.24564596}{doi:10.5281/zenodo.24564596}; its checksums are supplied with the reproduction material.
The public project is \url{https://github.com/wuwphy/RGZWorkflow}.
Historical versions 1.0.0 and 1.1.0 retain their own version-specific
DOIs~\cite{WuRGZWorkflow,WuRGZWorkflow110}; those identifiers do not
identify the corrected 1.1.2 release.
The unchanged scalar-integral catalogue permits reuse of numerical
master values on the same continued sheet; every corrected coefficient
and virtual interference is reevaluated. This is not described as
fresh quadrature of all scalar functions. The source-bound certificate,
coefficient comparison, and numerical input identities are distinguished
from the independent color-algebra checks and scalar-integral tests.

\appendix
\section{Quadratic kernels and ordered free fields}
\label{app:kernels}
Project the auxiliary color pair onto the normalized antisymmetric
adjoint channel. With $z=p^2$, $k=z-\mu^2$, and $h_G=m_G^2/2$, the
transverse field order is $(A,\phi,\bar\phi,\beta,\bar\beta)$ and
\begin{equation}
 K_T=\begin{pmatrix}
 m^2-z&h_G&h_G&-h_G&-h_G\\
 h_G&0&k&0&0\\ h_G&k&0&0&0\\
 -h_G&0&0&0&k\\-h_G&0&0&k&0
 \end{pmatrix}.
\end{equation}
Its $AA$ inverse gives Eq.~\eqref{eq:DAA}. The longitudinal order is
$(A_L,b,\tau,\xi,\phi_L,\bar\phi_L,\beta_L,\bar\beta_L)$.
Writing $q_z=\sqrt z$, $\bm h=h_G(1,1,-1,-1)$, and
$C=k\,\operatorname{diag}(\sigma_1,\sigma_1)$, the block form is
\begin{align}
 K_L={}&\begin{pmatrix}B&H\\\widetilde H&C\end{pmatrix},\quad
 H=\begin{pmatrix}\bm h\\\bm0\\\bm0\\-\ii q_z\bm h\end{pmatrix},
 \notag\\
 \widetilde H={}&\begin{pmatrix}\bm h^{\mathrm t}&\bm0^{\mathrm t}
 &\bm0^{\mathrm t}&\ii q_z\bm h^{\mathrm t}\end{pmatrix},
\end{align}
where the displayed $\widetilde H$ denotes the corresponding four
columns and
\begin{equation}
 B=\begin{pmatrix}
 m^2&\ii q_z&\ii q_z&\ii q_zm^2\\
 -\ii q_z&\alpha&0&0\\
 -\ii q_z&0&0&z\\
 -\ii q_zm^2&0&z&zm^2-M^4
 \end{pmatrix}.
\end{equation}
The oriented longitudinal modes at $p$ and $-p$ fix the off-diagonal
signs. The $\tau$ convention corresponds to
$-\partial\tau\cdot A^h$ in Eq.~\eqref{eq:Ltotal}.

The additional diagonal sectors provide independent checks of the
free action. For the ordered contraction
$\langle\chi(p)\bar\chi(-p)\rangle$,
\begin{align}
 D_{\eta\bar\eta}^{ab}&=\frac{\ii\delta^{ab}}{p^2},\notag\\
 D_{\omega\bar\omega,\mu\nu}^{ac,bd}
 &=D_{\zeta\bar\zeta,\mu\nu}^{ac,bd}
 =\frac{-\ii g_{\mu\nu}\delta^{ab}\delta^{cd}}{p^2-\mu^2},\notag\\
 S_f(p)&=\frac{\ii\slashp p}{p^2}\delta_{\rm color}.
\end{align}
Reversing a Grassmann field order includes its relative minus sign.
These formulae follow from the actual quadratic action, independently
of the generic propagator representation used to generate diagrams.

\section{Required expansion of the composite field}
\label{app:Ah}
The three-field expansion used to obtain all required vertices is
\begin{align}
 A_\mu^{h,a}={}&A_\mu^a-\partial_\mu\xi^a
 -g_sf^{abc}\left(A_\mu^b\xi^c
 +\tfrac12\xi^b\partial_\mu\xi^c\right)\notag\\
 &+\tfrac{g_s^2}{2}f^{abc}f^{cde}\xi^b\xi^dA_\mu^e
 \notag\\
 &-\tfrac{g_s^2}{6}f^{abc}f^{cde}
 \xi^b\xi^d\partial_\mu\xi^e+\cdots.
\end{align}
The omitted terms have at least four fields. They cannot enter the
one-loop amplitude after the quark-attachment argument of
Sec.~\ref{sec:construction}. Ghost four-point tadpoles vanish by
odd loop momentum rather than by the absence of those vertices.

\section{Reduction and reproducibility checks}
The action was implemented in FeynRules \cite{Alloul2014}, graph
amplitudes were generated with FeynArts \cite{Hahn2001}, and the
$D$-dimensional algebra and tensor reduction used FeynCalc
\cite{Shtabovenko2020,Shtabovenko2024}. Separate checks compare the
quadratic action with explicit Gaussian/Clifford inverses and compare
vertices under labeled leg permutations. The auxiliary matrix
connections and a directly generated quark self-energy provide
checks of momentum orientation and external residues.

At the scalar boundary, the full $D$ dependence of each reduction
coefficient is retained until the finite transfer is included. The
electronic export then checks all 144 finite-function coefficients
against the full coefficients of the source representation before
a row prefactor is removed. No numerical point is used for that
reconstruction or for Eq.~\eqref{eq:gaugezero}.

The counts of generators and integral entries should be interpreted
within the stated representation. Critical-point counting
\cite{Lee2013} and cut separation provide checks within the declared
generic-$D$ integral family; they do not prove a smallest possible
four-dimensional helicity representation or exclude identities at
exceptional mass values. Details of the electronic reconstruction
and integral argument order are supplied separately.

\subsection{Color and coefficient checks}
The current-kernel color identities include
\begin{equation}
 \sum_{b,c,d,e,h}f^{gbc}f^{bed}f^{eg'h}f^{hcd}
 =\frac{C_A^2}{2}\delta^{gg'}.
\end{equation}
They are checked before the corrected diagrams are reduced. The exact
export independently reconstructs every finite-function coefficient
from the complete corrected row. The comparison with the prior table
and its finite scale slope are supplied as executable checks. Historical
v1.1.0 example records remain identified as historical; they are not
substituted for this corrected-source evidence.

\section{Corrected evidence and reproducibility}
The corrected calculation is identified by its own bare source,
MSbar, general-gauge comparison, completed mass ledger and finite
audit. All 24 conditional certificate checks pass. The exact coefficient
export checks all 144 functions against the corrected factored rows;
14 full coefficients in $C_8\mathcal S_1$ differ from the previous
export. Neither historical certification labels nor an old software
version DOI are used to certify the new result.

The numerical update reuses scalar-integral values only after exact
identity of the entire 31-function catalogue, including argument order,
is established. All 34 saved kinematic/mass points are reevaluated with
the corrected coefficients, and energy, infrared and QCD-boundary
summaries are rebuilt. The scalar quadrature checks themselves do not
depend on the color contraction. New coefficient-scale checks pass
for all eight rows, as does the unchanged singlet spectral identity.
The supplied provenance distinguishes this algebraic/numerical update
from an independent repetition of every graph and every quadrature.
The corrected source snapshot is separately identified by its archive
hash; it is not silently equated with a mutable repository branch.

\section*{Acknowledgments and tool disclosure}
OpenAI Codex assisted code editing, manuscript preparation and
validation-script construction. AI assistance is not independent
physical verification; the author retains scientific responsibility.

\bibliographystyle{apsrev4-2}
\bibliography{references}

@article{Gribov1978,
  author = {V. N. Gribov},
  title = {{Quantization of non-Abelian gauge theories}},
  year = {1978},
  journal = {Nucl. Phys. B},
  volume = {139},
  pages = {1},
  doi = {10.1016/0550-3213(78)90175-X}
}

@article{Zwanziger1989,
  author = {D. Zwanziger},
  title = {{Local and renormalizable action from the Gribov horizon}},
  year = {1989},
  journal = {Nucl. Phys. B},
  volume = {323},
  pages = {513},
  doi = {10.1016/0550-3213(89)90122-3}
}

@article{Dudal2008,
  author = {D. Dudal and J. A. Gracey and S. P. Sorella and N. Vandersickel and H. Verschelde},
  title = {{A refinement of the Gribov-Zwanziger approach in the Landau gauge: Infrared propagators in harmony with the lattice results}},
  year = {2008},
  journal = {Phys. Rev. D},
  volume = {78},
  pages = {065047},
  doi = {10.1103/PhysRevD.78.065047},
  eprint = {0806.4348},
  archivePrefix = {arXiv}
}

@article{Capri2016,
  author = {M. A. L. Capri and D. Dudal and D. Fiorentini and M. S. Guimaraes and I. F. Justo and A. D. Pereira and B. W. Mintz and L. F. Palhares and R. F. Sobreiro and S. P. Sorella},
  title = {{A local and BRST-invariant Yang-Mills theory within the Gribov horizon}},
  year = {2016},
  journal = {Phys. Rev. D},
  volume = {94},
  pages = {025035},
  doi = {10.1103/PhysRevD.94.025035},
  eprint = {1605.02610},
  archivePrefix = {arXiv}
}

@article{Capri2017,
  author = {M. A. L. Capri and D. Fiorentini and A. D. Pereira and S. P. Sorella},
  title = {{Renormalizability of the Refined Gribov-Zwanziger action in the linear covariant gauges}},
  year = {2017},
  journal = {Phys. Rev. D},
  volume = {96},
  pages = {054022},
  doi = {10.1103/PhysRevD.96.054022},
  eprint = {1708.01543},
  archivePrefix = {arXiv}
}

@article{deBrito2025,
  author = {G. P. de Brito and P. De Fabritiis and A. D. Pereira},
  title = {{Quark propagator at one-loop in the Refined Gribov-Zwanziger framework}},
  year = {2025},
  journal = {Phys. Rev. D},
  volume = {112},
  pages = {014014},
  doi = {10.1103/lds6-34v8},
  eprint = {2505.10602},
  archivePrefix = {arXiv}
}

@article{Binosi2009,
  author = {D. Binosi and J. Papavassiliou},
  title = {{Pinch Technique: Theory and Applications}},
  year = {2009},
  journal = {Phys. Rept.},
  volume = {479},
  pages = {1--152},
  eprint = {0909.2536},
  archivePrefix = {arXiv}
}

@article{Passarino1979,
  author = {G. Passarino and M. J. G. Veltman},
  title = {{One-loop corrections for electron-positron annihilation into muon pairs in the Weinberg model}},
  year = {1979},
  journal = {Nucl. Phys. B},
  volume = {160},
  pages = {151--207},
  doi = {10.1016/0550-3213(79)90234-7}
}

@article{Denner2011,
  author = {A. Denner and S. Dittmaier},
  title = {{Scalar one-loop 4-point integrals}},
  year = {2011},
  journal = {Nucl. Phys. B},
  volume = {844},
  pages = {199--242},
  doi = {10.1016/j.nuclphysb.2010.11.002},
  eprint = {1005.2076},
  archivePrefix = {arXiv}
}

@article{Lee2013,
  author = {R. N. Lee and A. A. Pomeransky},
  title = {{Critical points and number of master integrals}},
  year = {2013},
  journal = {JHEP},
  number = {11},
  pages = {165},
  doi = {10.1007/JHEP11(2013)165},
  eprint = {1308.6676},
  archivePrefix = {arXiv}
}

@article{Alloul2014,
  author = {A. Alloul and N. D. Christensen and C. Degrande and C. Duhr and B. Fuks},
  title = {{FeynRules 2.0: A complete toolbox for tree-level phenomenology}},
  year = {2014},
  journal = {Comput. Phys. Commun.},
  volume = {185},
  pages = {2250--2300},
  doi = {10.1016/j.cpc.2014.04.012},
  eprint = {1310.1921},
  archivePrefix = {arXiv}
}

@article{Hahn2001,
  author = {T. Hahn},
  title = {{Generating Feynman diagrams and amplitudes with FeynArts 3}},
  year = {2001},
  journal = {Comput. Phys. Commun.},
  volume = {140},
  pages = {418--431},
  doi = {10.1016/S0010-4655(01)00290-9},
  eprint = {hep-ph/0012260},
  archivePrefix = {arXiv}
}

@article{Shtabovenko2020,
  author = {V. Shtabovenko and R. Mertig and F. Orellana},
  title = {{FeynCalc 9.3: New features and improvements}},
  year = {2020},
  journal = {Comput. Phys. Commun.},
  volume = {256},
  pages = {107478},
  doi = {10.1016/j.cpc.2020.107478},
  eprint = {2001.04407},
  archivePrefix = {arXiv}
}

@article{Shtabovenko2024,
  author = {V. Shtabovenko and R. Mertig and F. Orellana},
  title = {{FeynCalc 10: Do multiloop integrals dream of computer codes?}},
  year = {2025},
  journal = {Comput. Phys. Commun.},
  volume = {306},
  pages = {109357},
  doi = {10.1016/j.cpc.2024.109357},
  eprint = {2312.14089},
  archivePrefix = {arXiv}
}

@article{Patel2015,
  author = {H. H. Patel},
  title = {{Package-X: A Mathematica package for the analytic calculation of one-loop integrals}},
  year = {2015},
  journal = {Comput. Phys. Commun.},
  volume = {197},
  pages = {276--290},
  doi = {10.1016/j.cpc.2015.08.017},
  eprint = {1503.01469},
  archivePrefix = {arXiv}
}

@article{Shtabovenko2017,
  author = {V. Shtabovenko},
  title = {{FeynHelpers: Connecting FeynCalc to FIRE and Package-X}},
  year = {2017},
  journal = {Comput. Phys. Commun.},
  volume = {218},
  pages = {48--65},
  doi = {10.1016/j.cpc.2017.04.014},
  eprint = {1611.06793},
  archivePrefix = {arXiv}
}

@misc{WuRGZWorkflow,
  author = {Wan Wu},
  title = {{RGZWorkflow}},
  year = {2026},
  doi = {10.5281/zenodo.23724368},
  howpublished = {Version 1.0.0, open-source software},
  url = {https://github.com/wuwphy/RGZWorkflow/releases/tag/v1.0.0},
  note = {Released 28 October 2026; GPL-3.0-or-later}
}

@misc{WuRGZWorkflow110,
 author = {Wan Wu},
 title = {{RGZWorkflow}},
 year = {2026},
 doi = {10.5281/zenodo.24212586},
 howpublished = {Version 1.1.0, open-source software},
 url = {https://github.com/wuwphy/RGZWorkflow/releases/tag/v1.1.0},
 note = {Compact author validation record dated 14 November 2026; GPL-3.0-or-later}
}

@article{KelleySchwartz2011,
 author={Randall Kelley and Matthew D. Schwartz},
 title={1-loop matching and NNLL resummation for all partonic 2 to 2 processes in QCD},
 journal={Phys. Rev. D}, volume={83}, pages={045022}, year={2011},
 doi={10.1103/PhysRevD.83.045022}, eprint={1008.2759}, archivePrefix={arXiv}
}

@article{Feucht2004,
 author={B. Feucht and J. H. K{\"u}hn and A. A. Penin and V. A. Smirnov},
 title={Two-Loop Sudakov Form Factor in a Theory with a Mass Gap},
 journal={Phys. Rev. Lett.},volume={93},pages={101802},year={2004},
 doi={10.1103/PhysRevLett.93.101802},eprint={hep-ph/0404082},archivePrefix={arXiv}
}

@misc{SupplementalMaterial,
 year={2026},
 note={See Supplemental Material at [URL will be inserted by publisher] for the exact coefficient catalogue, numerical continuation, benchmark data and reproducibility instructions}
}

@misc{FeynCalcSUNFFix,
 author={FeynCalc developers},
 title={Correction of colorSimplifyGeneric},
 year={2026},
 howpublished={Source commit 1c5d2d88e9bd08738732486e45b97240ceca808c},
 url={https://github.com/FeynCalc/feyncalc/commit/1c5d2d88e9bd08738732486e45b97240ceca808c},
 note={3 September 2026; distinct index-order contraction patterns}
}

@misc{WuRGZWorkflow112,
 author={Wan Wu},
 title={{RGZWorkflow}, version 1.1.2: color-corrected source snapshot},
 year={2026},
 howpublished={Zenodo, GPL-3.0-or-later},
 doi={10.5281/zenodo.24564596},
 url={https://github.com/wuwphy/RGZWorkflow},
 note={Version 1.1.2; corrected release archive SHA-256 recorded in the reproduction material}
}
\clearpage
\onecolumngrid
\setcounter{section}{0}
\setcounter{equation}{0}
\setcounter{table}{0}
\renewcommand{\thesection}{S\arabic{section}}
\renewcommand{\theequation}{S\arabic{section}.\arabic{equation}}
\renewcommand{\thetable}{S\arabic{section}.\arabic{table}}
\renewcommand{\theHsection}{supp.\arabic{section}}
\renewcommand{\theHequation}{supp.\arabic{section}.\arabic{equation}}
\renewcommand{\theHtable}{supp.\arabic{section}.\arabic{table}}
\begin{center}
{\large\bfseries Supplemental Material}\\[6pt]
{\bfseries Gauge-parameter cancellation and finite form factors in one-loop
quark scattering in the refined Gribov--Zwanziger framework}\\[6pt]
Wan Wu --- Independent Researcher\\
\texttt{wuw20phy@gmail.com}
\end{center}
Electronic files are under \path{anc/reproduction/}; script and data paths below are relative to that directory.

\section{Exact electronic result and conventions}
This supplement specifies the finite one-loop amplitude, excluding the Born contribution. The accompanying Wolfram Language file \texttt{data/form-factors.wl} contains the authoritative exact coefficients. It requires no FeynCalc installation to read. Its symbols belong to the inert context \texttt{RGZPaper`}. The symbol \texttt{Finite} denotes the finite scalar integral defined in the main text, not an automatically evaluated function.

Write $a=r_-$, $b=r_+$, $u=-s-t$, and let \texttt{mu2} denote $\mu^2$, not $\mu_R^2$. All masses appearing as scalar-integral arguments are squared masses. With the color and spin structures of the main text,
\begin{equation}
\mathcal M^{(1)}_{\rm fin}=\sum_{c\in\{1,8\}}\sum_{i=1}^{4}C_c S_i P_{ci}\sum_{j=0}^{30} c_{ci,j}\,\mathcal F_j.
\end{equation}
ColorID 1 means $C_1=\delta_{31}\delta_{42}$; ColorID 2 means $C_8=T^a_{31}T^a_{42}$, rather than the crossed Kronecker tensor. SpinID labels $S_1,\ldots,S_4$ in the main text. Unlisted coefficients vanish. The eight rows contain 144 nonzero entries in total. These coefficients retain the common loop normalization in $P_{ci}$; do not multiply by an additional $1/(16\pi^2)$.

In the electronic file a function index $j$ points to \texttt{Functions[[j+1]]}; index zero denotes the rational remainder. Each row supplies \texttt{Prefactor} and \texttt{ReducedCoefficients}, the latter as pairs $\{j,c_{ci,j}\}$. Thus a row is reconstructed by multiplying its prefactor by the sum of each coefficient times its indexed function. The readable listing below restarts the notation \texttt{c[j]} for each row. It is a presentation of separate rows, not a single sequence of global assignments to execute.

\section{Additional checks supplied with the revision}
The coefficient table below is regenerated from the color-corrected
source. Fourteen full coefficients in the octet first-spin row change;
the other seven rows are exactly unchanged. The new-minus-old terms
are supplied in \texttt{data/colour-delta.wl}. The revised main article adds
an exact scale-response test on all eight rows, a spectral
reconstruction of the three singlet triangle terms, a conditional
coalescent-pole limit, and an explicit four-dimensional Born-interference
projection. The scripts in \texttt{tools/} and their results in
\texttt{checks/} accompany the revised LaTeX source. These scripts test
specified identities; they do not numerically evaluate all finite
multiscale functions or establish arbitrary-$N_c$ graph completeness.

\section{Color-correction and source binding}
The corrected export passes 144 exact coefficient reconstructions and
is linked to a new conditional certificate with all 24 checks passed.
The finite difference has the form $C_8\mathcal S_1\Delta F_{81}(t)$.
Its subtraction-scale slope vanishes exactly; no triangle or box
coefficient changes. The function indices affected are
$0,1,2,3,6,7,8,9,10,11,12,13,14,15$.
The correction file contains full coefficients: no additional row
prefactor should be multiplied into those differences.

The exact scalar catalogue is unchanged. The 34 numerical points
reuse the previously evaluated scalar functions on the same branch,
with the corrected coefficients inserted anew. The supplied checks
identify this reuse explicitly. Old graph-check records and old
software DOIs are historical and do not replace the corrected source
and certificate hashes.

\section{Integral arguments and branch prescription}
For chords $Q_0=0,Q_1,\ldots$, use denominators $(q+Q_j)^2-r_j+i0$. The three invariants of $C_0$ are chord distances $(01,12,02)$; the six invariants of $D_0$ are $(01,12,23,03,02,13)$. In particular,
\begin{align}
C_0 &: (Q_1^2,(Q_1-Q_2)^2,Q_2^2;r_0,r_1,r_2),\\
D_0 &: (Q_1^2,(Q_1-Q_2)^2,(Q_2-Q_3)^2,Q_3^2,Q_2^2,(Q_1-Q_3)^2;r_0,r_1,r_2,r_3).
\end{align}
Zero entries in these sequences are invariants, not omitted arguments. A simultaneous permutation of denominators and their masses may change the displayed invariant order. The bar denotes the finite Laurent coefficient with the subtraction and dimensional measure of the main text. All functions share the scale $\mu_R^2$ even where it is suppressed.

The square root defining $r_\pm$ and all logarithms are continued from the specified Euclidean parameter domain without a singularity crossing. For physical timelike invariants the indicated boundary value must be retained. The printed logarithms below suppress this boundary-value notation for brevity. Logarithm differences must not be combined using identities that change branches. The expression applies at nonexceptional invariants and nonzero, distinct pole masses; coalescent and zero-mass limits require separate limiting calculations.

\section{Finite-function catalogue}
The thirty-entry unreduced scalar inventory contains three tadpoles, twelve bubbles, nine triangles, and six boxes. After explicit tadpole and massless-bubble evaluation the finite representation uses the following 31 functions, including the rational unit and six logarithms. This inventory is not an assertion of minimal algebraic independence on every parameter locus.
\begin{center}
\begin{minipage}[t]{0.48\textwidth}
\centering
\begin{tabular}{rl}
\toprule $j$ & $\mathcal F_j$\\\midrule
0 & $1$ \\ 
1 & $\ln\!\left(\frac{r_-}{\mu _R^2}\right)$ \\ 
2 & $\ln\!\left(\frac{r_+}{\mu _R^2}\right)$ \\ 
3 & $\ln\!\left(\frac{\mu ^2}{\mu _R^2}\right)$ \\ 
4 & $\ln\!\left(-\frac{s}{\mu _R^2}\right)$ \\ 
5 & $\ln\!\left(\frac{s+t}{\mu _R^2}\right)$ \\ 
6 & $\ln\!\left(-\frac{t}{\mu _R^2}\right)$ \\ 
7 & $\overline{B}_0\!\left(t;0,r_-\right)$ \\ 
8 & $\overline{B}_0\!\left(t;0,r_+\right)$ \\ 
9 & $\overline{B}_0\!\left(t;0,\mu ^2\right)$ \\ 
10 & $\overline{B}_0\!\left(t;r_-,r_-\right)$ \\ 
11 & $\overline{B}_0\!\left(t;r_-,r_+\right)$ \\ 
12 & $\overline{B}_0\!\left(t;r_+,r_+\right)$ \\ 
13 & $\overline{B}_0\!\left(t;\mu ^2,r_-\right)$ \\ 
14 & $\overline{B}_0\!\left(t;\mu ^2,r_+\right)$ \\ 
15 & $\overline{B}_0\!\left(t;\mu ^2,\mu ^2\right)$ \\ 
\bottomrule
\end{tabular}
\end{minipage}
\hfill
\begin{minipage}[t]{0.48\textwidth}
\centering
\begin{tabular}{rl}
\toprule $j$ & $\mathcal F_j$\\\midrule
16 & $\overline{C}_0\!\left(0,t,0;0,r_-,r_-\right)$ \\ 
17 & $\overline{C}_0\!\left(0,t,0;0,r_-,r_+\right)$ \\ 
18 & $\overline{C}_0\!\left(0,t,0;0,r_+,r_+\right)$ \\ 
19 & $\overline{C}_0\!\left(s,0,0;0,0,r_-\right)$ \\ 
20 & $\overline{C}_0\!\left(s,0,0;0,0,r_+\right)$ \\ 
21 & $\overline{C}_0\!\left(-s-t,0,0;0,0,r_-\right)$ \\ 
22 & $\overline{C}_0\!\left(-s-t,0,0;0,0,r_+\right)$ \\ 
23 & $\overline{C}_0\!\left(t,0,0;0,0,r_-\right)$ \\ 
24 & $\overline{C}_0\!\left(t,0,0;0,0,r_+\right)$ \\ 
25 & $\overline{D}_0\!\left(s,0,t,0,0,0;0,0,r_-,r_-\right)$ \\ 
26 & $\overline{D}_0\!\left(s,0,t,0,0,0;0,0,r_-,r_+\right)$ \\ 
27 & $\overline{D}_0\!\left(s,0,t,0,0,0;0,0,r_+,r_+\right)$ \\ 
28 & $\overline{D}_0\!\left(-s-t,0,t,0,0,0;0,0,r_-,r_-\right)$ \\ 
29 & $\overline{D}_0\!\left(-s-t,0,t,0,0,0;0,0,r_-,r_+\right)$ \\ 
30 & $\overline{D}_0\!\left(-s-t,0,t,0,0,0;0,0,r_+,r_+\right)$ \\ 
\bottomrule
\end{tabular}
\end{minipage}
\end{center}
\section{Row prefactors}
In the following table $c=1,2$ refers to ColorID, with $c=2$ denoting the octet.
\begin{tabular}{ccr l}
\toprule ColorID & SpinID & Entries & $P_{ci}$\\\midrule
1 & 1 & 15 & $\frac{\left(N_c-1\right) \left(N_c+1\right) g_s^4}{16 \pi ^2 N_c^2 (s+t)}$ \\ 
2 & 1 & 31 & $\frac{g_s^4}{4 \pi ^2}$ \\ 
1 & 2 & 19 & $\frac{\left(N_c-1\right) \left(N_c+1\right) g_s^4}{8 \pi ^2 N_c^2 (s+t)}$ \\ 
2 & 2 & 19 & $\frac{g_s^4}{2 \pi ^2 t N_c (s+t)}$ \\ 
1 & 3 & 13 & $\frac{\left(N_c-1\right) \left(N_c+1\right) g_s^4}{64 \pi ^2 N_c^2 \left(r_--r_+\right)}$ \\ 
2 & 3 & 13 & $\frac{g_s^4}{32 \pi ^2 N_c \left(r_--r_+\right)}$ \\ 
1 & 4 & 15 & $\frac{\left(N_c-1\right) \left(N_c+1\right) g_s^4}{32 \pi ^2 s N_c^2 (s+t)}$ \\ 
2 & 4 & 19 & $\frac{g_s^4}{16 \pi ^2 s (s+t)}$ \\ 

\bottomrule\end{tabular}
\section{Numerical continuation and reproducibility}
The numerical additions reevaluate the corrected coefficient table using
the unchanged scalar-integral catalogue and its continued master values. On the
path $m^2=\mu^2=\lambda^2m_0^2$, $m_G^4=\lambda^4m_0^4/4$,
the scalar squared masses are $a,b=\lambda^2m_0^2(1\pm i/2)$.
We use $\log(-s-i0)=\log s-i\pi$, not a conjugation of the whole
integral when a mass is conjugated.

For the box family $D_0(v,0,t,0,0,0;0,0,a,b)$, fix two
Cheng--Wu parameters to sum to unity. Set
\begin{align}
 A&=ax+b(1-x),& B&=A-tx(1-x),\\
 C&=-v\xi(1-\xi)-i0,&d&=\sqrt{A^2-4BC}.
\end{align}
The remaining radial primitive is
\begin{equation}
 J(A,B,C)=\int_0^\infty\frac{r\,\dd r}{(Br^2+Ar+C)^2}
 =\frac{(A/d)[2\log(A+d)-\log(4B)-\log C]-2}{A^2-4BC}.
\end{equation}
Thus $D_0=\int_0^1\dd x\int_0^1\dd\xi\,J$. The separate logarithms
retain the continued sheet; merging them into a single principal log
can change the answer. At removable discriminant zeros a series is
used instead of subtracting nearly equal terms. The physical-region
integration is obtained from the Euclidean contour by rotating
$r=e^{-i\pi/4}\rho$. For $t<0$ and the mass phases used here,
$\operatorname{Im}(Ar)<0$ and $\operatorname{Im}(Br^2)<0$;
continuing $v$ to its physical value then avoids the radial poles.
The additional IR variations use $m_G^4/m_0^4=1/16,1/4,1$,
so the pole phases remain within $\lvert\arg a\rvert\leq\pi/4$.
This construction is restricted to these phases and does not assert
an unrestricted complex-mass integration rule.

Endpoint maps $x=\sin^2\varphi$ and
$\xi=(1-\cos\theta)/2$, together with $\theta=(\pi/2)e^{-z}$,
are used for stable quadrature. Working precision is 40 digits;
requested absolute and relative integration goals are 10 digits.
The $z$ integration is truncated at 36. These internal settings
are not a claim of 40 reliable output digits. The independent full
benchmark checks all six boxes with a different integration order
and the rotated radial contour. Small-mass scans also contain slow
convergence warnings; the QCD comparison is reported with fit-window
variation and an explicit finite-mass residual.

The electronic material includes:
\begin{itemize}
\item \texttt{tools/numerical\_masters.wl}, the restricted-family evaluator;
\item \texttt{tools/evaluate\_virtual.wls}, the complete finite-amplitude scan;
\item \texttt{data/numerical-benchmarks.wxf} and the CSV energy,
form-factor and QCD comparison tables;
\item independent parameter/contour checks and full-benchmark comparisons
in \texttt{checks/};
\item \texttt{tools/check\_qcd\_boundary.wls}, preserving $\epsilon$
before the massless triangle limit, and the four-helicity comparison
and free-intercept extrapolation scripts.
\end{itemize}
Run scripts from this directory with WolframKernel; set
\texttt{RGZ\_NUMERIC\_POINTS} to a Wolfram list of
$\{s/m_0^2,\lambda\}$ pairs to choose scan points. Optional
third and fourth entries choose $-t/s$ and $m_G^4/(\lambda^4m_0^4)$.

The extended IR comparison and angular/mass variation tables are
\begin{quote}\small
\path{data/infrared-comparison.csv}\\
\path{data/infrared-variations.csv}
\end{quote}
The script \path{tools/analyze_infrared.wls} checks exact scale
cancellation before plotting these combinations.

Numerical checkpoints are computational conveniences, distinct from
production model certificates. Existing benchmark records and exact
coefficients are retained; no coefficient was forced to a QCD value.

\section{Non-symmetric QCD comparison}
The additional boundary calculation uses $s=4m_0^2$,
$-t/s=1/3,2/3$ and five common masses from $\lambda=1/16$ to
$1/256$. The analysis, summary and comparison table are
\begin{quote}\small
\path{tools/compare_qcd_angles.wls}\\
\path{checks/qcd-angles-summary.json}\\
\path{data/qcd-angles-comparison.csv}
\end{quote}
They retain the actual complex residuals and free-intercept fits,
including all twelve limiting color--helicity coefficients for
three angles. These checks use the same QCD reference cited in
the main text and preserve distinct $t$ and $u$; they are not a
general analytic matching proof.

\section{Color-algebra validation and revision comparison}
\label{sec:colour-correction}
The coefficients in this revision incorporate a correction to the
structure-constant reduction used in the earlier calculation.
Define $F_2(b,a,c,d)=\sum_x f^{xba}f^{xcd}$. The identities
\begin{align}
 \sum_{a,c} F_2(b,a,c,d)f^{ace}&=-\frac{C_A}{2}f^{bde},\notag\\
 \sum_{a,c} F_2(b,c,a,d)f^{ace}&=+\frac{C_A}{2}f^{bde}
 \label{eq:colour-correction}
\end{align}
require distinct index-order patterns. A duplicated pattern in the
color-reduction implementation can overwrite the first contraction
and give an incorrect sign for larger networks. We use the corrected
rule, check the relevant color identities before calculation, and
identify the upstream correction explicitly~\cite{FeynCalcSUNFFix}.
This is a correction to algebraic reduction, not to the local action
or its free-field matrix.

A new source-bound calculation of the complete amplitude was performed
with the corrected rules. Its conditional certificate passes all
24 checks, including symbolic gauge cancellation, UV/IR closure,
the mass/scaleless ledger, finite assembly, and current-source binding.
The exact electronic result was re-exported and all 144 coefficient
reconstruction checks pass. Comparison with the previous table gives
\begin{equation}
 \Delta\mathcal M^{(1)}_{\rm fin}=C_8\mathcal S_1\Delta F_{81}(t),
 \quad\Delta F_{81}=\sum_{j\in\mathcal J}\Delta f_j\mathcal F_j,
 \label{eq:colour-delta}
\end{equation}
where
\begin{equation}
 \mathcal J=\{0,1,2,3,6,7,8,9,10,11,12,13,14,15\}.
\end{equation}
The differences here are full coefficients, including their row
prefactor. All seven other color--spin rows are unchanged exactly.
The integral catalogue is unchanged; only rational terms, logarithms,
and bubbles enter the difference. The machine-readable
\path{data/colour-delta.wl} specifies every $\Delta f_j$.

The correction is independent of $s$ and its exact subtraction-scale
slope is zero. It changes the finite amplitude and the virtual
Born interference, while leaving the previously tested scale response
and helicity differences that eliminate $\mathcal S_1$ unchanged.
Gauge independence does not imply that the finite amplitude is
unchanged: the corrected full calculation retains vanishing gauge
difference, while Eq.~\eqref{eq:colour-delta} is nonzero.
All coefficient tables and numerical virtual results in this revision
use the corrected result. The conditional analytic domain and the
limitations concerning colored asymptotic states are retained.

\section{Corrected source and historical checks}
The new source-bound certificate passes 24 declared-domain checks,
and the exact export passes 144 coefficient reconstruction comparisons.
The comparison file identifies 14 changed full coefficients, all in
the octet first-spin row. Its zero scale slope is checked exactly.
The archived v1.1.0 example checks are historical; they must not be
used to identify the corrected source snapshot or numerical tables.
The corrected certificate, source hashes, electronic coefficient
comparison and numerical refresh records accompany this revision.

\section{Provenance and scope}
The implementation snapshot is RGZWorkflow 1.1.2, GPL-3.0-or-later.
Its distributed archive SHA-256 is the concatenation of
\begin{quote}\scriptsize\ttfamily
db8e0d84fdf503cd01486f2a91ac4c06\\
d781867e9af58a0d3ba0662251da8d76
\end{quote}
The two lines concatenate.
The corrected release DOI is \href{https://doi.org/10.5281/zenodo.24564596}{10.5281/zenodo.24564596}.
This hash identifies the current release package, excluding the additional
PT research material included in the initial package. Historical version
DOIs and the initial package hash do not identify this release archive.
The exact result export has its own calculation-source fingerprint
and hashes for the final scalar rows, MSbar, scalar inventory and
conditional certificate. All 144 coefficients were reconstructed
exactly after multiplication by their row prefactors. This establishes
faithful export, distinct from independent verification of every graph.

The new full calculation retains the stated regulator limit,
nonzero-mass IR conditions and common-sheet continuation. Numerical
updates reuse scalar values only after exact equality of every integral
argument and catalogue entry is verified, and then insert the corrected
coefficients. No inclusive cross section or unrestricted physical
asymptotic-state claim is made.

\section{All reduced coefficients}
For convenience, the symbols used in the coefficient listing are
\begin{align}
 a=r_-&=\frac{m^2+\mu^2}{2}
 -i\sqrt{m_G^4-\frac{(m^2-\mu^2)^2}{4}},\\
 b=r_+&=\frac{m^2+\mu^2}{2}
 +i\sqrt{m_G^4-\frac{(m^2-\mu^2)^2}{4}}.
\end{align}
Both $a$ and $b$ are squared pole masses, with the square-root branch
specified in the main text. In the listing, \texttt{mu2} means $\mu^2$,
\texttt{gs} means $g_s$, \texttt{Nc} means $N_c$, and \texttt{s},
\texttt{t} are the Mandelstam invariants. The coefficient \texttt{c[j]}
multiplies $\mathcal F_j$ in the catalogue above; \texttt{c[0]} is the
rational remainder. Each color--spin block has its own coefficients and
common \texttt{Prefactor}. The displayed representation assumes $a\ne b$;
coincident poles require a separate limit.

\begin{Verbatim}[fontsize=\scriptsize]
(* Color 1; spin 1. F = Prefactor * Sum[c[j] Function[j]]. *)
Prefactor = (gs^4*(-1 + Nc)*(1 + Nc))/(16*Nc^2*Pi^2*(s + t));
c[4] = s/t;
c[5] = -(s/t);
c[16] = -1/2*((-a + mu2)^2*(2*a - t)*(s^2 + t^2))/((a - b)^2*s*(s + t));
c[17] = ((-a + mu2)*(-b + mu2)*(a + b - t)*(s^2 + t^2))/((a - b)^2*s*(s + t));
c[18] = -1/2*((-b + mu2)^2*(2*b - t)*(s^2 + t^2))/((-a + b)^2*s*(s + t));
c[19] = -(((-a + mu2)*s*(-(b*s^2) + mu2*s^2 - 2*b*s*t + 2*mu2*s*t + 2*a*t^2 - t^3))/((a - b)*t^2*(s + t)));
c[20] = ((-b + mu2)*s*(-(a*s^2) + mu2*s^2 - 2*a*s*t + 2*mu2*s*t + 2*b*t^2 - t^3))/((a - b)*t^2*(s + t));
c[21] = (-2*(-a + mu2)*(-b + mu2)*s^3 - (-a + mu2)*s^2*(-3*b + 3*mu2 - t)*t + 2*(-a + mu2)*s*t^2*(a + t) + (-a + mu2)*t^3*(-2*a
- b + mu2 + t))/(2*(a - b)*s*t^2);
c[22] = -1/2*(-2*(-a + mu2)*(-b + mu2)*s^3 - (-b + mu2)*s^2*(-3*a + 3*mu2 - t)*t + 2*(-b + mu2)*s*t^2*(b + t) + (-b +
mu2)*t^3*(-a - 2*b + mu2 + t))/((a - b)*s*t^2);
c[25] = ((-a + mu2)^2*s*(2*a^2*s + 2*a*s^2 + 6*a^2*t + 4*a*s*t - s^2*t - 2*a*t^2 - 2*s*t^2))/(2*(a - b)^2*t*(s + t));
c[26] = (-((-a + mu2)*(-b + mu2)*s*t^2*(6*a*b - a*t - b*t)) + 2*(-a + mu2)*(-b + mu2)*s^2*t*(a^2 - 3*a*b + b^2 - a*t - b*t +
t^2) + (-a + mu2)*(-b + mu2)*s^3*(a^2 - 2*a*b + b^2 - a*t - b*t + t^2))/((a - b)^2*t^2*(s + t));
c[27] = ((-b + mu2)^2*s*(2*b^2*s + 2*b*s^2 + 6*b^2*t + 4*b*s*t - s^2*t - 2*b*t^2 - 2*s*t^2))/(2*(-a + b)^2*t*(s + t));
c[28] = -1/4*((-a + mu2)^2*(4*a^2*s^2 - 4*a*s^3 + 4*a^2*s*t - 4*a*s^2*t + 2*s^3*t - 4*a^2*t^2 + 4*a*s*t^2 + 3*s^2*t^2 + 4*a*t^3
- t^4))/((a - b)^2*s*t);
c[29] = (-((-a + mu2)*(-b + mu2)*(a + b - t)^2*t^3) + 2*(-a + mu2)*(-b + mu2)*s*t^2*(2*a*b + a*t + b*t) + 2*(-a + mu2)*(-b +
mu2)*s^3*(a^2 - 2*a*b + b^2 - a*t - b*t + t^2) + (-a + mu2)*(-b + mu2)*s^2*t*(3*a^2 - 2*a*b + 3*b^2 - 2*a*t - 2*b*t +
3*t^2))/(2*(a - b)^2*s*t^2);
c[30] = -1/4*((-b + mu2)^2*(4*b^2*s^2 - 4*b*s^3 + 4*b^2*s*t - 4*b*s^2*t + 2*s^3*t - 4*b^2*t^2 + 4*b*s*t^2 + 3*s^2*t^2 + 4*b*t^3
- t^4))/((-a + b)^2*s*t);

\end{Verbatim}

\begin{Verbatim}[fontsize=\scriptsize]
(* Color 2; spin 1. F = Prefactor * Sum[c[j] Function[j]]. *)
Prefactor = gs^4/(4*Pi^2);
c[0] = (3*a^2*b^2*Nc^2*t^6*(s + t) + 6*a*b*mu2*Nc^2*t^5*(s + t)*(a*b - a*t - b*t) + 3*mu2^6*Nc^2*(s + t)*(154*a^2*b^2 + 9*t^4) -
6*mu2^5*Nc^2*(s + t)*(86*a^3*b^2 + 86*a^2*b^3 + 124*a^2*b^2*t - 12*a*b*t^3 + 9*a*t^4 + 9*b*t^4 + 3*t^5) + mu2^4*(s +
t)*(-1152*a^3*b^3 + 54*a^4*b^2*Nc^2 - 576*a^3*b^3*Nc^2 + 54*a^2*b^4*Nc^2 + 1152*a^3*b^2*t + 1152*a^2*b^3*t + 2088*a^3*b^2*Nc^2*t
+ 2088*a^2*b^3*Nc^2*t - 1152*a^2*b^2*t^2 + 1280*a^2*b^2*Nc*t^2 - 1408*a^2*b^2*Nc^2*t^2 - 72*a^2*b*Nc^2*t^3 - 72*a*b^2*Nc^2*t^3 +
27*a^2*Nc^2*t^4 + 12*a*b*Nc^2*t^4 + 27*b^2*Nc^2*t^4 + 36*a*Nc^2*t^5 + 36*b*Nc^2*t^5 + 3*Nc^2*t^6) + 2*mu2^3*(s + t)*(576*a^4*b^3
+ 576*a^3*b^4 + 546*a^4*b^3*Nc^2 + 546*a^3*b^4*Nc^2 - 576*a^4*b^2*t + 432*a^3*b^3*t - 576*a^2*b^4*t - 672*a^4*b^2*Nc^2*t +
84*a^3*b^3*Nc^2*t - 672*a^2*b^4*Nc^2*t - 1008*a^3*b^2*t^2 - 1008*a^2*b^3*t^2 - 1224*a^3*b^2*Nc^2*t^2 - 1224*a^2*b^3*Nc^2*t^2 +
1584*a^2*b^2*t^3 - 1280*a^2*b^2*Nc*t^3 + 2092*a^2*b^2*Nc^2*t^3 + 21*a^2*b*Nc^2*t^4 + 21*a*b^2*Nc^2*t^4 - 9*a^2*Nc^2*t^5 -
24*a*b*Nc^2*t^5 - 9*b^2*Nc^2*t^5 - 3*a*Nc^2*t^6 - 3*b*Nc^2*t^6) + mu2^2*(2310*a^4*b^4*Nc^2*s + 6*a^4*b^4*Nc^2*t -
1152*a^4*b^3*s*t - 1152*a^3*b^4*s*t - 5568*a^4*b^3*Nc^2*s*t - 5568*a^3*b^4*Nc^2*s*t - 1152*a^4*b^3*t^2 - 1152*a^3*b^4*t^2 -
960*a^4*b^3*Nc^2*t^2 - 960*a^3*b^4*Nc^2*t^2 + 1152*a^4*b^2*s*t^2 + 288*a^3*b^3*s*t^2 + 1152*a^2*b^4*s*t^2 +
3456*a^4*b^2*Nc^2*s*t^2 + 9072*a^3*b^3*Nc^2*s*t^2 + 3456*a^2*b^4*Nc^2*s*t^2 + 1152*a^4*b^2*t^3 + 288*a^3*b^3*t^3 +
1152*a^2*b^4*t^3 + 1152*a^4*b^2*Nc^2*t^3 - 144*a^3*b^3*Nc^2*t^3 + 1152*a^2*b^4*Nc^2*t^3 + 864*a^3*b^2*s*t^3 + 864*a^2*b^3*s*t^3
- 3312*a^3*b^2*Nc^2*s*t^3 - 3312*a^2*b^3*Nc^2*s*t^3 + 864*a^3*b^2*t^4 + 864*a^2*b^3*t^4 + 1296*a^3*b^2*Nc^2*t^4 +
1296*a^2*b^3*Nc^2*t^4 - 2016*a^2*b^2*s*t^4 + 1280*a^2*b^2*Nc*s*t^4 - 469*a^2*b^2*Nc^2*s*t^4 - 2016*a^2*b^2*t^5 +
1280*a^2*b^2*Nc*t^5 - 2773*a^2*b^2*Nc^2*t^5 + 12*a^2*b*Nc^2*s*t^5 + 12*a*b^2*Nc^2*s*t^5 + 12*a^2*b*Nc^2*t^6 + 12*a*b^2*Nc^2*t^6
+ 3*a^2*Nc^2*s*t^6 + 12*a*b*Nc^2*s*t^6 + 3*b^2*Nc^2*s*t^6 + 3*a^2*Nc^2*t^7 + 12*a*b*Nc^2*t^7 +
3*b^2*Nc^2*t^7))/(2304*a^2*b^2*mu2^2*Nc*(a - t)^2*(b - t)^2*t*(s + t));
c[3] = (-((-a + mu2)*(-b + mu2)*(a^2 + a*b + b^2 - 2*a*mu2 - 2*b*mu2 + 13*mu2^2)*Nc) - (-5*a^2*b - 5*a*b^2 + 5*a^2*mu2 -
38*a*b*mu2 + 5*b^2*mu2 + 43*a*mu2^2 + 43*b*mu2^2)*Nc*t + 2*(-5*a*b + 5*a*mu2 + 5*b*mu2 + 43*mu2^2)*Nc*t^2 -
48*mu2*Nc*t^3)/(768*(a - t)^2*t*(-b + t)^2);
c[4] = -(s/(Nc*t*(s + t)));
c[6] = (768*a^3*b^3*mu2^3*(-a - b + mu2) - 384*a^2*b^2*mu2^2*(-2*a^2*b - 2*a*b^2 - 2*a^2*mu2 + a*b*mu2 - 2*b^2*mu2 + 2*a*mu2^2 +
2*b*mu2^2)*t + 16*a^2*b^2*mu2^2*(-48*a^2 - 24*a*b - 48*b^2 + 72*a*mu2 + 72*b*mu2 + 48*mu2^2 - 32*mu2^2*Nc + mu2^2*Nc^2)*t^2 -
8*a*b*mu2^2*(48*a^2*b + 48*a*b^2 + 240*a*b*mu2 - 128*a*b*mu2*Nc + 7*a*b*mu2*Nc^2 - 3*a*mu2^2*Nc^2 - 3*b*mu2^2*Nc^2 +
3*mu2^3*Nc^2)*t^3 - mu2^2*(-1152*a^2*b^2 + 512*a^2*b^2*Nc - 39*a^2*b^2*Nc^2 + 14*a^2*b*mu2*Nc^2 + 14*a*b^2*mu2*Nc^2 +
9*a^2*mu2^2*Nc^2 + 4*a*b*mu2^2*Nc^2 + 9*b^2*mu2^2*Nc^2 - 18*a*mu2^3*Nc^2 - 18*b*mu2^3*Nc^2 + 9*mu2^4*Nc^2)*t^4 + 2*mu2*(-a +
mu2)*(-b + mu2)*(-(a*b) - 3*a*mu2 - 3*b*mu2 + 3*mu2^2)*Nc^2*t^5 - (-a + mu2)^2*(-b + mu2)^2*Nc^2*t^6)/(1536*a^2*b^2*mu2^2*Nc*(a
- t)^2*(b - t)^2*t);
c[5] = -1/2*((-2 + Nc^2)*s)/(Nc*t*(s + t));
c[1] = (-192*(-a + mu2)*(a - t)*(b - t)*(mu2 - t)*(2*a + 3*t) + Nc^2*(-6*a^3*b^2 - 4*a^2*b^3 - a*b^4 - 356*a^3*b*mu2 +
28*a^2*b^2*mu2 + 11*a*b^3*mu2 + b^4*mu2 - 22*a^3*mu2^2 + 188*a^2*b*mu2^2 - 37*a*b^2*mu2^2 - 7*b^3*mu2^2 + 172*a^2*mu2^3 +
189*a*b*mu2^3 + 15*b^2*mu2^3 - 162*a*mu2^4 - 21*b*mu2^4 + 12*mu2^5 + 320*a^3*b*t - 4*a^2*b^2*t + 5*a*b^3*t + 448*a^3*mu2*t -
216*a^2*b*mu2*t - 13*a*b^2*mu2*t - 5*b^3*mu2*t - 980*a^2*mu2^2*t - 293*a*b*mu2^2*t + 17*b^2*mu2^2*t + 733*a*mu2^3*t +
141*b*mu2^3*t - 153*mu2^4*t - 384*a^3*t^2 + 88*a^2*b*t^2 - 10*a*b^2*t^2 + 1160*a^2*mu2*t^2 + 772*a*b*mu2*t^2 + 10*b^2*mu2*t^2 -
1050*a*mu2^2*t^2 - 764*b*mu2^2*t^2 + 178*mu2^3*t^2 - 432*a^2*t^3 - 640*a*b*t^3 - 80*a*mu2*t^3 + 592*b*mu2*t^3 + 560*mu2^2*t^3 +
576*a*t^4 - 576*mu2*t^4))/(768*(a - b)*Nc*(a - t)^2*(b - t)^2*t);
c[2] = (192*(-b + mu2)*(a - t)*(b - t)*(mu2 - t)*(2*b + 3*t) + Nc^2*(a^4*b + 4*a^3*b^2 + 6*a^2*b^3 - a^4*mu2 - 11*a^3*b*mu2 -
28*a^2*b^2*mu2 + 356*a*b^3*mu2 + 7*a^3*mu2^2 + 37*a^2*b*mu2^2 - 188*a*b^2*mu2^2 + 22*b^3*mu2^2 - 15*a^2*mu2^3 - 189*a*b*mu2^3 -
172*b^2*mu2^3 + 21*a*mu2^4 + 162*b*mu2^4 - 12*mu2^5 - 5*a^3*b*t + 4*a^2*b^2*t - 320*a*b^3*t + 5*a^3*mu2*t + 13*a^2*b*mu2*t +
216*a*b^2*mu2*t - 448*b^3*mu2*t - 17*a^2*mu2^2*t + 293*a*b*mu2^2*t + 980*b^2*mu2^2*t - 141*a*mu2^3*t - 733*b*mu2^3*t +
153*mu2^4*t + 10*a^2*b*t^2 - 88*a*b^2*t^2 + 384*b^3*t^2 - 10*a^2*mu2*t^2 - 772*a*b*mu2*t^2 - 1160*b^2*mu2*t^2 + 764*a*mu2^2*t^2
+ 1050*b*mu2^2*t^2 - 178*mu2^3*t^2 + 640*a*b*t^3 + 432*b^2*t^3 - 592*a*mu2*t^3 + 80*b*mu2*t^3 - 560*mu2^2*t^3 - 576*b*t^4 +
576*mu2*t^4))/(768*(a - b)*Nc*(a - t)^2*(b - t)^2*t);
c[9] = -1/768*((-a + mu2)*(-b + mu2)*Nc*(mu2 - t)^2*(12*mu2^4 - 60*mu2^3*t + 13*mu2^2*t^2 - 2*mu2*t^3 + t^4))/(a*b*mu2^2*(a -
t)^2*(b - t)^2*t);
c[7] = ((-a + mu2)*(-b + mu2)^2*Nc*(a^4 - 6*a^3*mu2 + 9*a^2*mu2^2 - 4*a^3*t + 6*a^2*mu2*t + 90*a*mu2^2*t + 6*a^2*t^2 +
6*a*mu2*t^2 + 9*mu2^2*t^2 - 4*a*t^3 - 6*mu2*t^3 + t^4))/(768*a^2*(a - b)*b*mu2*(b - t)^2*t);
c[8] = -1/768*((-a + mu2)^2*(-b + mu2)*Nc*(b^4 - 6*b^3*mu2 + 9*b^2*mu2^2 - 4*b^3*t + 6*b^2*mu2*t + 90*b*mu2^2*t + 6*b^2*t^2 +
6*b*mu2*t^2 + 9*mu2^2*t^2 - 4*b*t^3 - 6*mu2*t^3 + t^4))/(a*(a - b)*b^2*mu2*(a - t)^2*t);
c[15] = (Nc*(mu2 - t)^2*(4*mu2 - t)*(84*mu2^2 + 4*mu2*t - t^2))/(1536*mu2^2*(-a + t)^2*(-b + t)^2);
c[13] = -1/768*((-b + mu2)*Nc*(a^2 - 2*a*mu2 + mu2^2 - 2*a*t - 2*mu2*t + t^2)*(a^4 - 2*a^3*mu2 + 13*a^2*mu2^2 - 60*a*mu2^3 +
12*mu2^4 - 4*a^3*t - 46*a^2*mu2*t + 178*a*mu2^2*t - 60*mu2^3*t + 6*a^2*t^2 - 46*a*mu2*t^2 + 13*mu2^2*t^2 - 4*a*t^3 - 2*mu2*t^3 +
t^4))/(a*(a - b)*mu2*(a - t)^2*(b - t)^2*t);
c[14] = ((-a + mu2)*Nc*(b^2 - 2*b*mu2 + mu2^2 - 2*b*t - 2*mu2*t + t^2)*(b^4 - 2*b^3*mu2 + 13*b^2*mu2^2 - 60*b*mu2^3 + 12*mu2^4 -
4*b^3*t - 46*b^2*mu2*t + 178*b*mu2^2*t - 60*mu2^3*t + 6*b^2*t^2 - 46*b*mu2*t^2 + 13*mu2^2*t^2 - 4*b*t^3 - 2*mu2*t^3 +
t^4))/(768*(a - b)*b*mu2*(a - t)^2*(b - t)^2*t);
c[10] = ((-b + mu2)^2*Nc*(3*mu2 - t)^2*t^4*(s + t) - 768*a^6*Nc*(b - t)*(b*s + mu2*s + mu2*t - 2*s*t - t^2) + 4*a*(-b +
mu2)*Nc*t^3*(s + t)*(-42*b*mu2^2 + 36*mu2^3 - b*mu2*t + 9*mu2^2*t + b*t^2 - 3*mu2*t^2) + 64*a^5*Nc*(27*b^2*mu2*s + 21*b*mu2^2*s
+ 3*b^2*mu2*t + 21*b*mu2^2*t + 20*b^2*s*t - 68*b*mu2*s*t - 33*mu2^2*s*t - 4*b^2*t^2 - 20*b*mu2*t^2 - 33*mu2^2*t^2 - 47*b*s*t^2 +
65*mu2*s*t^2 + b*t^3 + 41*mu2*t^3 + 15*s*t^3 - 9*t^4) - 16*a^3*Nc*t*(-111*b^2*mu2^2*s + 24*b*mu2^3*s + 27*mu2^4*s -
15*b^2*mu2^2*t + 24*b*mu2^3*t + 27*mu2^4*t - 113*b^2*mu2*s*t + 239*b*mu2^2*s*t - 132*mu2^3*s*t - 17*b^2*mu2*t^2 + 47*b*mu2^2*t^2
- 132*mu2^3*t^2 + 3*b^2*s*t^2 + 181*b*mu2*s*t^2 + 8*mu2^2*s*t^2 + 3*b^2*t^3 - 11*b*mu2*t^3 + 104*mu2^2*t^3 - 2*b*s*t^3 -
124*mu2*s*t^3 - 2*b*t^4 - 28*mu2*t^4) - 4*a^2*Nc*t^2*(273*b^2*mu2^2*s - 174*b*mu2^3*s + 153*mu2^4*s + 81*b^2*mu2^2*t -
174*b*mu2^3*t + 153*mu2^4*t + 58*b^2*mu2*s*t - 490*b*mu2^2*s*t - 72*mu2^3*s*t + 58*b^2*mu2*t^2 - 106*b*mu2^2*t^2 - 72*mu2^3*t^2
- 8*b^2*s*t^2 + 16*b*mu2*s*t^2 + 244*mu2^2*s*t^2 - 8*b^2*t^3 + 16*b*mu2*t^3 + 52*mu2^2*t^3 + 2*b*s*t^3 - 2*mu2*s*t^3 + 2*b*t^4 -
2*mu2*t^4) - 16*a^4*Nc*(60*b^2*mu2^2*s + 36*b*mu2^3*s + 12*b^2*mu2^2*t + 36*b*mu2^3*t + 194*b^2*mu2*s*t - 158*b*mu2^2*s*t -
108*mu2^3*s*t + 2*b^2*mu2*t^2 - 62*b*mu2^2*t^2 - 108*mu2^3*t^2 + 37*b^2*s*t^2 - 368*b*mu2*s*t^2 + 250*mu2^2*s*t^2 - 11*b^2*t^3 +
16*b*mu2*t^3 + 202*mu2^2*t^3 - 92*b*s*t^3 + 86*mu2*s*t^3 + 4*b*t^4 - 106*mu2*t^4 + 63*s*t^4 + 15*t^5))/(1536*a^2*(a - b)^2*(a -
t)^2*(b - t)^2*t*(s + t));
c[11] = (-((-a + mu2)*(-b + mu2)*(a^6 + 2*a^5*b - a^4*b^2 - 772*a^3*b^3 - a^2*b^4 + 2*a*b^5 + b^6 - 6*a^5*mu2 - 6*a^4*b*mu2 -
276*a^3*b^2*mu2 - 276*a^2*b^3*mu2 - 6*a*b^4*mu2 - 6*b^5*mu2 + 9*a^4*mu2^2 + 72*a^3*b*mu2^2 - 162*a^2*b^2*mu2^2 + 72*a*b^3*mu2^2
+ 9*b^4*mu2^2)*Nc*s) - (-a + mu2)*(-b + mu2)*Nc*(a^6 + 2*a^5*b - a^4*b^2 - 4*a^3*b^3 - a^2*b^4 + 2*a*b^5 + b^6 - 6*a^5*mu2 -
6*a^4*b*mu2 - 276*a^3*b^2*mu2 - 276*a^2*b^3*mu2 - 6*a*b^4*mu2 - 6*b^5*mu2 + 9*a^4*mu2^2 + 72*a^3*b*mu2^2 - 162*a^2*b^2*mu2^2 +
72*a*b^3*mu2^2 + 9*b^4*mu2^2 - 6*a^5*s + 10*a^4*b*s + 1788*a^3*b^2*s + 1788*a^2*b^3*s + 10*a*b^4*s - 6*b^5*s + 18*a^4*mu2*s +
48*a^3*b*mu2*s + 348*a^2*b^2*mu2*s + 48*a*b^3*mu2*s + 18*b^4*mu2*s + 72*a^3*mu2^2*s - 288*a^2*b*mu2^2*s - 288*a*b^2*mu2^2*s +
72*b^3*mu2^2*s)*t - (-a + mu2)*(-b + mu2)*Nc*(-6*a^5 + 10*a^4*b + 252*a^3*b^2 + 252*a^2*b^3 + 10*a*b^4 - 6*b^5 + 18*a^4*mu2 +
48*a^3*b*mu2 + 348*a^2*b^2*mu2 + 48*a*b^3*mu2 + 18*b^4*mu2 + 72*a^3*mu2^2 - 288*a^2*b*mu2^2 - 288*a*b^2*mu2^2 + 72*b^3*mu2^2 +
15*a^4*s - 1012*a^3*b*s - 3142*a^2*b^2*s - 1012*a*b^3*s + 15*b^4*s - 12*a^3*mu2*s + 804*a^2*b*mu2*s + 804*a*b^2*mu2*s -
12*b^3*mu2*s - 162*a^2*mu2^2*s - 288*a*b*mu2^2*s - 162*b^2*mu2^2*s)*t^2 + (-a + mu2)*(-b + mu2)*Nc*(-15*a^4 + 244*a^3*b +
70*a^2*b^2 + 244*a*b^3 - 15*b^4 + 12*a^3*mu2 - 804*a^2*b*mu2 - 804*a*b^2*mu2 + 12*b^3*mu2 + 162*a^2*mu2^2 + 288*a*b*mu2^2 +
162*b^2*mu2^2 + 20*a^3*s - 1212*a^2*b*s - 1212*a*b^2*s + 20*b^3*s + 12*a^2*mu2*s - 288*a*b*mu2*s + 12*b^2*mu2*s - 72*a*mu2^2*s -
72*b*mu2^2*s)*t^3 - (-a + mu2)*(-b + mu2)*Nc*(-20*a^3 - 324*a^2*b - 324*a*b^2 - 20*b^3 - 12*a^2*mu2 + 288*a*b*mu2 - 12*b^2*mu2 +
72*a*mu2^2 + 72*b*mu2^2 + 15*a^2*s - 974*a*b*s + 15*b^2*s + 18*a*mu2*s + 18*b*mu2*s + 9*mu2^2*s)*t^4 - (-a + mu2)*(-b +
mu2)*Nc*(15*a^2 - 206*a*b + 15*b^2 + 18*a*mu2 + 18*b*mu2 + 9*mu2^2 - 6*a*s - 6*b*s - 6*mu2*s)*t^5 + (-a + mu2)*(-b +
mu2)*Nc*(6*a + 6*b + 6*mu2 - s)*t^6 - (-a + mu2)*(-b + mu2)*Nc*t^7)/(768*a*(a - b)^2*b*(a - t)^2*(b - t)^2*t*(s + t));
c[12] = ((-a + mu2)^2*Nc*(3*mu2 - t)^2*t^4*(s + t) - 768*b^6*Nc*(a - t)*(a*s + mu2*s + mu2*t - 2*s*t - t^2) + 4*b*(-a +
mu2)*Nc*t^3*(s + t)*(-42*a*mu2^2 + 36*mu2^3 - a*mu2*t + 9*mu2^2*t + a*t^2 - 3*mu2*t^2) + 64*b^5*Nc*(27*a^2*mu2*s + 21*a*mu2^2*s
+ 3*a^2*mu2*t + 21*a*mu2^2*t + 20*a^2*s*t - 68*a*mu2*s*t - 33*mu2^2*s*t - 4*a^2*t^2 - 20*a*mu2*t^2 - 33*mu2^2*t^2 - 47*a*s*t^2 +
65*mu2*s*t^2 + a*t^3 + 41*mu2*t^3 + 15*s*t^3 - 9*t^4) - 16*b^3*Nc*t*(-111*a^2*mu2^2*s + 24*a*mu2^3*s + 27*mu2^4*s -
15*a^2*mu2^2*t + 24*a*mu2^3*t + 27*mu2^4*t - 113*a^2*mu2*s*t + 239*a*mu2^2*s*t - 132*mu2^3*s*t - 17*a^2*mu2*t^2 + 47*a*mu2^2*t^2
- 132*mu2^3*t^2 + 3*a^2*s*t^2 + 181*a*mu2*s*t^2 + 8*mu2^2*s*t^2 + 3*a^2*t^3 - 11*a*mu2*t^3 + 104*mu2^2*t^3 - 2*a*s*t^3 -
124*mu2*s*t^3 - 2*a*t^4 - 28*mu2*t^4) - 4*b^2*Nc*t^2*(273*a^2*mu2^2*s - 174*a*mu2^3*s + 153*mu2^4*s + 81*a^2*mu2^2*t -
174*a*mu2^3*t + 153*mu2^4*t + 58*a^2*mu2*s*t - 490*a*mu2^2*s*t - 72*mu2^3*s*t + 58*a^2*mu2*t^2 - 106*a*mu2^2*t^2 - 72*mu2^3*t^2
- 8*a^2*s*t^2 + 16*a*mu2*s*t^2 + 244*mu2^2*s*t^2 - 8*a^2*t^3 + 16*a*mu2*t^3 + 52*mu2^2*t^3 + 2*a*s*t^3 - 2*mu2*s*t^3 + 2*a*t^4 -
2*mu2*t^4) - 16*b^4*Nc*(60*a^2*mu2^2*s + 36*a*mu2^3*s + 12*a^2*mu2^2*t + 36*a*mu2^3*t + 194*a^2*mu2*s*t - 158*a*mu2^2*s*t -
108*mu2^3*s*t + 2*a^2*mu2*t^2 - 62*a*mu2^2*t^2 - 108*mu2^3*t^2 + 37*a^2*s*t^2 - 368*a*mu2*s*t^2 + 250*mu2^2*s*t^2 - 11*a^2*t^3 +
16*a*mu2*t^3 + 202*mu2^2*t^3 - 92*a*s*t^3 + 86*mu2*s*t^3 + 4*a*t^4 - 106*mu2*t^4 + 63*s*t^4 + 15*t^5))/(1536*(a - b)^2*b^2*(a -
t)^2*(b - t)^2*t*(s + t));
c[16] = (4*(-a + mu2)^2*(a - t)*(2*a - t)*(b - t)*t*(s^2 + t^2) + (-a + mu2)*Nc^2*(s + t)^2*(-8*a^3*b*s - 4*a^3*mu2*s +
9*a^2*b*mu2*s + 3*a^2*mu2^2*s + 4*a^3*b*t - 4*a^2*b*mu2*t + 12*a^3*s*t + 9*a^2*b*s*t - 17*a^2*mu2*s*t - 10*a*b*mu2*s*t +
6*a*mu2^2*s*t - 4*a^3*t^2 - 6*a^2*b*t^2 + 4*a^2*mu2*t^2 + 6*a*b*mu2*t^2 - 4*a^2*s*t^2 - 4*a*b*s*t^2 + 4*a*mu2*s*t^2 +
4*b*mu2*s*t^2 + 6*a^2*t^3 + 2*a*b*t^3 - 6*a*mu2*t^3 - 2*b*mu2*t^3 + 4*a*s*t^3 - 4*mu2*s*t^3 - 2*a*t^4 + 2*mu2*t^4))/(8*(a -
b)^2*Nc*s*(a - t)*(b - t)*t*(s + t)^2);
c[17] = (-4*(-a + mu2)*(-b + mu2)*(a - t)*(b - t)*(a + b - t)*t*(s^2 + t^2) - (-a + mu2)*(-b + mu2)*Nc^2*(s + t)^2*(4*a^2*b*s +
4*a*b^2*s + 3*a*b*mu2*s - 2*a^2*b*t - 2*a*b^2*t - 4*a^2*s*t - 13*a*b*s*t - 4*b^2*s*t + 3*a*mu2*s*t + 3*b*mu2*s*t + 2*a^2*t^2 +
6*a*b*t^2 + 2*b^2*t^2 + 4*a*s*t^2 + 4*b*s*t^2 - 4*a*t^3 - 4*b*t^3 - 4*s*t^3 + 2*t^4))/(4*(a - b)^2*Nc*s*(a - t)*(b - t)*t*(s +
t)^2);
c[18] = (4*(-b + mu2)^2*(a - t)*(b - t)*(2*b - t)*t*(s^2 + t^2) + (-b + mu2)*Nc^2*(s + t)^2*(-8*a*b^3*s + 9*a*b^2*mu2*s -
4*b^3*mu2*s + 3*b^2*mu2^2*s + 4*a*b^3*t - 4*a*b^2*mu2*t + 9*a*b^2*s*t + 12*b^3*s*t - 10*a*b*mu2*s*t - 17*b^2*mu2*s*t +
6*b*mu2^2*s*t - 6*a*b^2*t^2 - 4*b^3*t^2 + 6*a*b*mu2*t^2 + 4*b^2*mu2*t^2 - 4*a*b*s*t^2 - 4*b^2*s*t^2 + 4*a*mu2*s*t^2 +
4*b*mu2*s*t^2 + 2*a*b*t^3 + 6*b^2*t^3 - 2*a*mu2*t^3 - 6*b*mu2*t^3 + 4*b*s*t^3 - 4*mu2*s*t^3 - 2*b*t^4 + 2*mu2*t^4))/(8*(a -
b)^2*Nc*s*(a - t)*(b - t)*t*(s + t)^2);
c[19] = ((-a + mu2)*s*(-(b*s^2) + mu2*s^2 - 2*b*s*t + 2*mu2*s*t + 2*a*t^2 - t^3))/((a - b)*Nc*t^2*(s + t)^2);
c[20] = ((-b + mu2)*s*(-(a*s^2) + mu2*s^2 - 2*a*s*t + 2*mu2*s*t + 2*b*t^2 - t^3))/((-a + b)*Nc*t^2*(s + t)^2);
c[21] = -1/4*((-a + mu2)*(-2 + Nc^2)*(-2*b*s^3 + 2*mu2*s^3 - 3*b*s^2*t + 3*mu2*s^2*t - 2*a*s*t^2 - s^2*t^2 + 2*a*t^3 + b*t^3 -
mu2*t^3 - 2*s*t^3 - t^4))/((a - b)*Nc*s*t^2*(s + t));
c[22] = -1/4*((-b + mu2)*(-2 + Nc^2)*(-2*a*s^3 + 2*mu2*s^3 - 3*a*s^2*t + 3*mu2*s^2*t - 2*b*s*t^2 - s^2*t^2 + a*t^3 + 2*b*t^3 -
mu2*t^3 - 2*s*t^3 - t^4))/((-a + b)*Nc*s*t^2*(s + t));
c[23] = -1/2*((-a + mu2)*(mu2 - t)*(a + t)^2)/((a - b)*Nc*(a - t)*(b - t)*t);
c[24] = -1/2*((-b + mu2)*(mu2 - t)*(b + t)^2)/((-a + b)*Nc*(a - t)*(b - t)*t);
c[25] = -1/2*((-a + mu2)^2*s*(2*a^2*s + 2*a*s^2 + 6*a^2*t + 4*a*s*t - s^2*t - 2*a*t^2 - 2*s*t^2))/((a - b)^2*Nc*t*(s + t)^2);
c[26] = ((-a + mu2)*(-b + mu2)*s*t^2*(6*a*b - a*t - b*t) - 2*(-a + mu2)*(-b + mu2)*s^2*t*(a^2 - 3*a*b + b^2 - a*t - b*t + t^2) -
(-a + mu2)*(-b + mu2)*s^3*(a^2 - 2*a*b + b^2 - a*t - b*t + t^2))/((a - b)^2*Nc*t^2*(s + t)^2);
c[27] = -1/2*((-b + mu2)^2*s*(2*b^2*s + 2*b*s^2 + 6*b^2*t + 4*b*s*t - s^2*t - 2*b*t^2 - 2*s*t^2))/((-a + b)^2*Nc*t*(s + t)^2);
c[28] = -1/8*((-a + mu2)^2*(-2 + Nc^2)*(4*a^2*s^2 - 4*a*s^3 + 4*a^2*s*t - 4*a*s^2*t + 2*s^3*t - 4*a^2*t^2 + 4*a*s*t^2 +
3*s^2*t^2 + 4*a*t^3 - t^4))/((a - b)^2*Nc*s*t*(s + t));
c[29] = (-((-a + mu2)*(-b + mu2)*(-2 + Nc^2)*(a + b - t)^2*t^3) + 2*(-a + mu2)*(-b + mu2)*(-2 + Nc^2)*s*t^2*(2*a*b + a*t + b*t)
+ 2*(-a + mu2)*(-b + mu2)*(-2 + Nc^2)*s^3*(a^2 - 2*a*b + b^2 - a*t - b*t + t^2) + (-a + mu2)*(-b + mu2)*(-2 + Nc^2)*s^2*t*(3*a^2
- 2*a*b + 3*b^2 - 2*a*t - 2*b*t + 3*t^2))/(4*(a - b)^2*Nc*s*t^2*(s + t));
c[30] = -1/8*((-b + mu2)^2*(-2 + Nc^2)*(4*b^2*s^2 - 4*b*s^3 + 4*b^2*s*t - 4*b*s^2*t + 2*s^3*t - 4*b^2*t^2 + 4*b*s*t^2 +
3*s^2*t^2 + 4*b*t^3 - t^4))/((-a + b)^2*Nc*s*t*(s + t));

\end{Verbatim}

\begin{Verbatim}[fontsize=\scriptsize]
(* Color 1; spin 2. F = Prefactor * Sum[c[j] Function[j]]. *)
Prefactor = (gs^4*(-1 + Nc)*(1 + Nc))/(8*Nc^2*Pi^2*(s + t));
c[0] = -s^(-1);
c[4] = -t^(-1);
c[5] = (2*s + t)/(2*s*t);
c[10] = (-a + mu2)^2/(2*(a - b)^2*s);
c[11] = -(((-a + mu2)*(-b + mu2))/((a - b)^2*s));
c[12] = (-b + mu2)^2/(2*(-a + b)^2*s);
c[16] = ((-a + mu2)^2*(2*a*s^2 - 3*s^2*t + 2*a*t^2 - 2*s*t^2 - t^3))/(2*(a - b)^2*s^2*(s + t));
c[17] = -(((-a + mu2)*(-b + mu2)*(a*s^2 + b*s^2 - 3*s^2*t + a*t^2 + b*t^2 - 2*s*t^2 - t^3))/((a - b)^2*s^2*(s + t)));
c[18] = ((-b + mu2)^2*(2*b*s^2 - 3*s^2*t + 2*b*t^2 - 2*s*t^2 - t^3))/(2*(-a + b)^2*s^2*(s + t));
c[19] = ((-a + mu2)*(-(b*s^2) + mu2*s^2 - 2*b*s*t + 2*mu2*s*t + 2*a*t^2 + s*t^2))/((a - b)*t^2*(s + t));
c[20] = ((-b + mu2)*(-(a*s^2) + mu2*s^2 - 2*a*s*t + 2*mu2*s*t + 2*b*t^2 + s*t^2))/((-a + b)*t^2*(s + t));
c[21] = (2*(-a + mu2)*(-b + mu2)*s^3 + (-a + mu2)*s^2*(-3*b + 3*mu2 - t)*t - 2*(-a + mu2)*s*t^3 - (-a + mu2)*t^3*(-2*a - b + mu2
+ t))/(2*(a - b)*s^2*t^2);
c[22] = -1/2*(2*(-a + mu2)*(-b + mu2)*s^3 + (-b + mu2)*s^2*(-3*a + 3*mu2 - t)*t - 2*(-b + mu2)*s*t^3 - (-b + mu2)*t^3*(-a - 2*b
+ mu2 + t))/((a - b)*s^2*t^2);
c[25] = -1/2*((-a + mu2)^2*(2*a^2*s + 2*a*s^2 + 6*a^2*t + 6*a*s*t - s*t^2))/((a - b)^2*t*(s + t));
c[26] = (6*a*b*(-a + mu2)*(-b + mu2)*t^2 - (-a + mu2)*(-b + mu2)*s^2*(a^2 - 2*a*b + b^2 - a*t - b*t) - (-a + mu2)*(-b +
mu2)*s*t*(2*a^2 - 6*a*b + 2*b^2 - 3*a*t - 3*b*t + t^2))/((a - b)^2*t^2*(s + t));
c[27] = -1/2*((-b + mu2)^2*(2*b^2*s + 2*b*s^2 + 6*b^2*t + 6*b*s*t - s*t^2))/((-a + b)^2*t*(s + t));
c[28] = ((-a + mu2)^2*(4*a^2*s^2 - 4*a*s^3 + 2*a^2*s*t - 4*a*s^2*t - 4*a^2*t^2 + 4*a*s*t^2 - s^2*t^2 + 4*a*t^3 - 2*s*t^3 -
t^4))/(4*(a - b)^2*s^2*t);
c[29] = ((-a + mu2)*(-b + mu2)*(a + b - t)^2*t^3 - 2*(-a + mu2)*(-b + mu2)*s^3*(a^2 - 2*a*b + b^2 - a*t - b*t) - (-a + mu2)*(-b
+ mu2)*s^2*t*(3*a^2 - 2*a*b + 3*b^2 - 2*a*t - 2*b*t - t^2) - 2*(-a + mu2)*(-b + mu2)*s*t^2*(a*b + a*t + b*t - t^2))/(2*(a -
b)^2*s^2*t^2);
c[30] = ((-b + mu2)^2*(4*b^2*s^2 - 4*b*s^3 + 2*b^2*s*t - 4*b*s^2*t - 4*b^2*t^2 + 4*b*s*t^2 - s^2*t^2 + 4*b*t^3 - 2*s*t^3 -
t^4))/(4*(-a + b)^2*s^2*t);

\end{Verbatim}

\begin{Verbatim}[fontsize=\scriptsize]
(* Color 2; spin 2. F = Prefactor * Sum[c[j] Function[j]]. *)
Prefactor = gs^4/(2*Nc*Pi^2*t*(s + t));
c[0] = -1/2*(2*Nc^2*s - 2*t + Nc^2*t)/s;
c[4] = 1;
c[5] = ((-2 + Nc^2)*(2*s + t))/(4*s);
c[10] = ((-a + mu2)^2*(2*Nc^2*s - 2*t + Nc^2*t))/(4*(a - b)^2*s);
c[11] = -1/2*((-a + mu2)*(-b + mu2)*(2*Nc^2*s - 2*t + Nc^2*t))/((a - b)^2*s);
c[12] = ((-b + mu2)^2*(2*Nc^2*s - 2*t + Nc^2*t))/(4*(-a + b)^2*s);
c[16] = (-((-a + mu2)^2*Nc^2*(s + t)^2*(4*a*s - 2*a*t + t^2)) - 2*(-a + mu2)^2*t*(2*a*s^2 - 3*s^2*t + 2*a*t^2 - 2*s*t^2 -
t^3))/(4*(a - b)^2*s^2*(s + t));
c[17] = ((-a + mu2)*(-b + mu2)*Nc^2*(s + t)^2*(2*a*s + 2*b*s - a*t - b*t + t^2) + 2*(-a + mu2)*(-b + mu2)*t*(a*s^2 + b*s^2 -
3*s^2*t + a*t^2 + b*t^2 - 2*s*t^2 - t^3))/(2*(a - b)^2*s^2*(s + t));
c[18] = (-((-b + mu2)^2*Nc^2*(s + t)^2*(4*b*s - 2*b*t + t^2)) - 2*(-b + mu2)^2*t*(2*b*s^2 - 3*s^2*t + 2*b*t^2 - 2*s*t^2 -
t^3))/(4*(a - b)^2*s^2*(s + t));
c[19] = -(((-a + mu2)*(-(b*s^2) + mu2*s^2 - 2*b*s*t + 2*mu2*s*t + 2*a*t^2 + s*t^2))/((a - b)*t*(s + t)));
c[20] = ((-b + mu2)*(-(a*s^2) + mu2*s^2 - 2*a*s*t + 2*mu2*s*t + 2*b*t^2 + s*t^2))/((a - b)*t*(s + t));
c[21] = ((-a + mu2)*(-2 + Nc^2)*(-2*b*s^3 + 2*mu2*s^3 - 3*b*s^2*t + 3*mu2*s^2*t - s^2*t^2 + 2*a*t^3 + b*t^3 - mu2*t^3 - 2*s*t^3
- t^4))/(4*(a - b)*s^2*t);
c[22] = ((-b + mu2)*(-2 + Nc^2)*(-2*a*s^3 + 2*mu2*s^3 - 3*a*s^2*t + 3*mu2*s^2*t - s^2*t^2 + a*t^3 + 2*b*t^3 - mu2*t^3 - 2*s*t^3
- t^4))/(4*(-a + b)*s^2*t);
c[25] = ((-a + mu2)^2*(2*a^2*s + 2*a*s^2 + 6*a^2*t + 6*a*s*t - s*t^2))/(2*(a - b)^2*(s + t));
c[26] = (-6*a*b*(-a + mu2)*(-b + mu2)*t^2 + (-a + mu2)*(-b + mu2)*s^2*(a^2 - 2*a*b + b^2 - a*t - b*t) + (-a + mu2)*(-b +
mu2)*s*t*(2*a^2 - 6*a*b + 2*b^2 - 3*a*t - 3*b*t + t^2))/((a - b)^2*t*(s + t));
c[27] = ((-b + mu2)^2*(2*b^2*s + 2*b*s^2 + 6*b^2*t + 6*b*s*t - s*t^2))/(2*(-a + b)^2*(s + t));
c[28] = ((-a + mu2)^2*(-2 + Nc^2)*(4*a^2*s^2 - 4*a*s^3 + 2*a^2*s*t - 4*a*s^2*t - 4*a^2*t^2 + 4*a*s*t^2 - s^2*t^2 + 4*a*t^3 -
2*s*t^3 - t^4))/(8*(a - b)^2*s^2);
c[29] = ((-a + mu2)*(-b + mu2)*(-2 + Nc^2)*(a + b - t)^2*t^3 - 2*(-a + mu2)*(-b + mu2)*(-2 + Nc^2)*s^3*(a^2 - 2*a*b + b^2 - a*t
- b*t) - (-a + mu2)*(-b + mu2)*(-2 + Nc^2)*s^2*t*(3*a^2 - 2*a*b + 3*b^2 - 2*a*t - 2*b*t - t^2) - 2*(-a + mu2)*(-b + mu2)*(-2 +
Nc^2)*s*t^2*(a*b + a*t + b*t - t^2))/(4*(a - b)^2*s^2*t);
c[30] = ((-b + mu2)^2*(-2 + Nc^2)*(4*b^2*s^2 - 4*b*s^3 + 2*b^2*s*t - 4*b*s^2*t - 4*b^2*t^2 + 4*b*s*t^2 - s^2*t^2 + 4*b*t^3 -
2*s*t^3 - t^4))/(8*(-a + b)^2*s^2);

\end{Verbatim}

\begin{Verbatim}[fontsize=\scriptsize]
(* Color 1; spin 3. F = Prefactor * Sum[c[j] Function[j]]. *)
Prefactor = (gs^4*(-1 + Nc)*(1 + Nc))/(64*(a - b)*Nc^2*Pi^2);
c[16] = ((-a + mu2)^2*(2*a - t)*t)/(2*(a - b)*s*(s + t));
c[17] = -(((-a + mu2)*(-b + mu2)*(a + b - t)*t)/((a - b)*s*(s + t)));
c[18] = -1/2*((-b + mu2)^2*(2*b - t)*t)/((-a + b)*s*(s + t));
c[19] = -1/2*((-a + mu2)*(-(b*s) + mu2*s + 2*a*t + s*t))/(t*(s + t));
c[20] = ((-b + mu2)*(-(a*s) + mu2*s + 2*b*t + s*t))/(2*t*(s + t));
c[21] = -1/2*((-a + mu2)*(-(b*s) + mu2*s - 2*a*t - b*t + mu2*t + s*t + t^2))/(s*t);
c[22] = ((-b + mu2)*(-(a*s) + mu2*s - a*t - 2*b*t + mu2*t + s*t + t^2))/(2*s*t);
c[25] = ((-a + mu2)^2*(4*a^2 + 4*a*s - s*t))/(4*(a - b)*(s + t));
c[26] = ((-a + mu2)*(-b + mu2)*(a^2*s - 2*a*b*s + b^2*s - 4*a*b*t - 2*a*s*t - 2*b*s*t + s*t^2))/(2*(a - b)*t*(s + t));
c[27] = -1/4*((-b + mu2)^2*(4*b^2 + 4*b*s - s*t))/((-a + b)*(s + t));
c[28] = -1/4*((-a + mu2)^2*(4*a^2 - 4*a*s - 4*a*t + s*t + t^2))/((a - b)*s);
c[29] = ((-a + mu2)*(-b + mu2)*(a + b - t)^2*t + (-a + mu2)*(-b + mu2)*s*(a^2 - 2*a*b + b^2 - 2*a*t - 2*b*t + t^2))/(2*(a -
b)*s*t);
c[30] = ((-b + mu2)^2*(4*b^2 - 4*b*s - 4*b*t + s*t + t^2))/(4*(-a + b)*s);

\end{Verbatim}

\begin{Verbatim}[fontsize=\scriptsize]
(* Color 2; spin 3. F = Prefactor * Sum[c[j] Function[j]]. *)
Prefactor = gs^4/(32*(a - b)*Nc*Pi^2);
c[16] = ((-a + mu2)^2*(2*a - t)*(Nc^2*s - 2*t + Nc^2*t))/(2*(a - b)*s*(s + t));
c[17] = -(((-a + mu2)*(-b + mu2)*(a + b - t)*(Nc^2*s - 2*t + Nc^2*t))/((a - b)*s*(s + t)));
c[18] = -1/2*((-b + mu2)^2*(2*b - t)*(Nc^2*s - 2*t + Nc^2*t))/((-a + b)*s*(s + t));
c[19] = ((-a + mu2)*(-(b*s) + mu2*s + 2*a*t + s*t))/(t*(s + t));
c[20] = -(((-b + mu2)*(-(a*s) + mu2*s + 2*b*t + s*t))/(t*(s + t)));
c[21] = -1/2*((-a + mu2)*(-2 + Nc^2)*(-(b*s) + mu2*s - 2*a*t - b*t + mu2*t + s*t + t^2))/(s*t);
c[22] = ((-b + mu2)*(-2 + Nc^2)*(-(a*s) + mu2*s - a*t - 2*b*t + mu2*t + s*t + t^2))/(2*s*t);
c[25] = -1/2*((-a + mu2)^2*(4*a^2 + 4*a*s - s*t))/((a - b)*(s + t));
c[26] = -(((-a + mu2)*(-b + mu2)*(a^2*s - 2*a*b*s + b^2*s - 4*a*b*t - 2*a*s*t - 2*b*s*t + s*t^2))/((a - b)*t*(s + t)));
c[27] = ((-b + mu2)^2*(4*b^2 + 4*b*s - s*t))/(2*(-a + b)*(s + t));
c[28] = -1/4*((-a + mu2)^2*(-2 + Nc^2)*(4*a^2 - 4*a*s - 4*a*t + s*t + t^2))/((a - b)*s);
c[29] = ((-a + mu2)*(-b + mu2)*(-2 + Nc^2)*(a + b - t)^2*t + (-a + mu2)*(-b + mu2)*(-2 + Nc^2)*s*(a^2 - 2*a*b + b^2 - 2*a*t -
2*b*t + t^2))/(2*(a - b)*s*t);
c[30] = ((-b + mu2)^2*(-2 + Nc^2)*(4*b^2 - 4*b*s - 4*b*t + s*t + t^2))/(4*(-a + b)*s);

\end{Verbatim}

\begin{Verbatim}[fontsize=\scriptsize]
(* Color 1; spin 4. F = Prefactor * Sum[c[j] Function[j]]. *)
Prefactor = (gs^4*(-1 + Nc)*(1 + Nc))/(32*Nc^2*Pi^2*s*(s + t));
c[4] = 1/2;
c[5] = -1/2;
c[16] = -1/2*((-a + mu2)^2*(2*a - t)*t*(2*s + t))/((a - b)^2*s*(s + t));
c[17] = ((-a + mu2)*(-b + mu2)*(a + b - t)*t*(2*s + t))/((a - b)^2*s*(s + t));
c[18] = -1/2*((-b + mu2)^2*(2*b - t)*t*(2*s + t))/((-a + b)^2*s*(s + t));
c[19] = ((-a + mu2)*(-(b*s^2) + mu2*s^2 + 4*a*s*t + s^2*t + 2*a*t^2))/(2*(a - b)*t*(s + t));
c[20] = ((-b + mu2)*(-(a*s^2) + mu2*s^2 + 4*b*s*t + s^2*t + 2*b*t^2))/(2*(-a + b)*t*(s + t));
c[21] = ((-a + mu2)*s^2*(-b + mu2 + t) + 2*(-a + mu2)*s*t*(-2*a - b + mu2 + t) + (-a + mu2)*t^2*(-2*a - b + mu2 + t))/(2*(a -
b)*s*t);
c[22] = -1/2*((-b + mu2)*s^2*(-a + mu2 + t) + 2*(-b + mu2)*s*t*(-a - 2*b + mu2 + t) + (-b + mu2)*t^2*(-a - 2*b + mu2 + t))/((a -
b)*s*t);
c[25] = -1/4*((-a + mu2)^2*(6*a^2*s + 4*a*s^2 + 2*a^2*t - s^2*t))/((a - b)^2*(s + t));
c[26] = -1/2*((-a + mu2)*(-b + mu2)*(a^2*s^2 - 2*a*b*s^2 + b^2*s^2 - 6*a*b*s*t - 2*a*s^2*t - 2*b*s^2*t - 2*a*b*t^2 +
s^2*t^2))/((a - b)^2*t*(s + t));
c[27] = -1/4*((-b + mu2)^2*(6*b^2*s + 4*b*s^2 + 2*b^2*t - s^2*t))/((-a + b)^2*(s + t));
c[28] = ((-a + mu2)^2*(6*a^2*s - 4*a*s^2 + 4*a^2*t - 8*a*s*t + s^2*t - 4*a*t^2 + 2*s*t^2 + t^3))/(4*(a - b)^2*s);
c[29] = (-((-a + mu2)*(-b + mu2)*(a + b - t)^2*t^2) - (-a + mu2)*(-b + mu2)*s^2*(a^2 - 2*a*b + b^2 - 2*a*t - 2*b*t + t^2) -
2*(-a + mu2)*(-b + mu2)*s*t*(a^2 + a*b + b^2 - 2*a*t - 2*b*t + t^2))/(2*(a - b)^2*s*t);
c[30] = ((-b + mu2)^2*(6*b^2*s - 4*b*s^2 + 4*b^2*t - 8*b*s*t + s^2*t - 4*b*t^2 + 2*s*t^2 + t^3))/(4*(-a + b)^2*s);

\end{Verbatim}

\begin{Verbatim}[fontsize=\scriptsize]
(* Color 2; spin 4. F = Prefactor * Sum[c[j] Function[j]]. *)
Prefactor = gs^4/(16*Pi^2*s*(s + t));
c[0] = Nc;
c[4] = -Nc^(-1);
c[5] = -1/2*(-2 + Nc^2)/Nc;
c[10] = -1/2*((-a + mu2)^2*Nc)/(a - b)^2;
c[11] = ((-a + mu2)*(-b + mu2)*Nc)/(a - b)^2;
c[12] = -1/2*((-b + mu2)^2*Nc)/(-a + b)^2;
c[16] = -1/2*((-a + mu2)^2*(2*a - t)*(Nc^2*s^2 - 4*s*t + 2*Nc^2*s*t - 2*t^2 + Nc^2*t^2))/((a - b)^2*Nc*s*(s + t));
c[17] = ((-a + mu2)*(-b + mu2)*(a + b - t)*(Nc^2*s^2 - 4*s*t + 2*Nc^2*s*t - 2*t^2 + Nc^2*t^2))/((a - b)^2*Nc*s*(s + t));
c[18] = -1/2*((-b + mu2)^2*(2*b - t)*(Nc^2*s^2 - 4*s*t + 2*Nc^2*s*t - 2*t^2 + Nc^2*t^2))/((-a + b)^2*Nc*s*(s + t));
c[19] = -(((-a + mu2)*(-(b*s^2) + mu2*s^2 + 4*a*s*t + s^2*t + 2*a*t^2))/((a - b)*Nc*t*(s + t)));
c[20] = ((-b + mu2)*(-(a*s^2) + mu2*s^2 + 4*b*s*t + s^2*t + 2*b*t^2))/((a - b)*Nc*t*(s + t));
c[21] = ((-a + mu2)*(-2 + Nc^2)*s^2*(-b + mu2 + t) + 2*(-a + mu2)*(-2 + Nc^2)*s*t*(-2*a - b + mu2 + t) + (-a + mu2)*(-2 +
Nc^2)*t^2*(-2*a - b + mu2 + t))/(2*(a - b)*Nc*s*t);
c[22] = -1/2*((-b + mu2)*(-2 + Nc^2)*s^2*(-a + mu2 + t) + 2*(-b + mu2)*(-2 + Nc^2)*s*t*(-a - 2*b + mu2 + t) + (-b + mu2)*(-2 +
Nc^2)*t^2*(-a - 2*b + mu2 + t))/((a - b)*Nc*s*t);
c[25] = ((-a + mu2)^2*(6*a^2*s + 4*a*s^2 + 2*a^2*t - s^2*t))/(2*(a - b)^2*Nc*(s + t));
c[26] = ((-a + mu2)*(-b + mu2)*(a^2*s^2 - 2*a*b*s^2 + b^2*s^2 - 6*a*b*s*t - 2*a*s^2*t - 2*b*s^2*t - 2*a*b*t^2 + s^2*t^2))/((a -
b)^2*Nc*t*(s + t));
c[27] = ((-b + mu2)^2*(6*b^2*s + 4*b*s^2 + 2*b^2*t - s^2*t))/(2*(-a + b)^2*Nc*(s + t));
c[28] = ((-a + mu2)^2*(-2 + Nc^2)*(6*a^2*s - 4*a*s^2 + 4*a^2*t - 8*a*s*t + s^2*t - 4*a*t^2 + 2*s*t^2 + t^3))/(4*(a - b)^2*Nc*s);
c[29] = (-((-a + mu2)*(-b + mu2)*(-2 + Nc^2)*(a + b - t)^2*t^2) - (-a + mu2)*(-b + mu2)*(-2 + Nc^2)*s^2*(a^2 - 2*a*b + b^2 -
2*a*t - 2*b*t + t^2) - 2*(-a + mu2)*(-b + mu2)*(-2 + Nc^2)*s*t*(a^2 + a*b + b^2 - 2*a*t - 2*b*t + t^2))/(2*(a - b)^2*Nc*s*t);
c[30] = ((-b + mu2)^2*(-2 + Nc^2)*(6*b^2*s - 4*b*s^2 + 4*b^2*t - 8*b*s*t + s^2*t - 4*b*t^2 + 2*s*t^2 + t^3))/(4*(-a +
b)^2*Nc*s);
\end{Verbatim}

\end{document}